\documentstyle[11pt,aaspp4]{article}
\begin{document}
\newcommand\ssalb{$\tilde\omega_0$}

\title{Reflected Spectra and Albedos of Extrasolar Giant Planets I:
Clear and Cloudy Atmospheres}
\author{Mark S. Marley, Christopher Gelino, Denise Stephens}
\affil{New Mexico State University}
\authoremail{mmarley@nmsu.edu}
\author{Jonathan I. Lunine}
\affil{Lunar and Planetary Laboratory, University of Arizona and
Reparto di Planetologia, Istituto di Astrophysica, Rome, Italy}
\author{Richard Freedman}
\affil{Space Physics Research Institute, NASA/Ames Research Center}

\begin{abstract}
The reflected spectra of extrasolar giant planets are primarily influenced
by Rayleigh scattering, molecular absorption, and atmospheric condensates.  
We present model geometric albedo and phase integral spectra and Bond
albedos for planets and brown dwarfs with masses between 0.8 and 70
Jupiter masses.
Rayleigh scattering predominates in the blue while molecular
absorption removes most red and infrared photons.  Thus cloud-free 
atmospheres, found on giant planets with effective
temperatures exceeding about 400 K, are quite dark in reflected light
beyond $0.6\,\rm \mu m$.   In cooler atmospheres first
water clouds and then other condensates provide a bright reflecting
layer.  Only planets with cloudy atmospheres will be detectable
in reflected light beyond $1 \, \rm \mu m$.  Thermal emission
dominates  the near-infrared for warm objects with clear atmospheres.
However the presence
of other condensates, not considered here, may brighten some planets
in reflected near-infrared light and darken them in the blue and UV.
Bond albedos, the ratio of the
total reflected to incident power, are sensitive to the spectral
type of the primary.  Most incident photons from early type stars
will be Rayleigh scattered, while most incident photons
from late type stars will be absorbed. 
The Bond albedo of a given planet thus may range from 0.4 to 0.05, depending
on the primary type.  Condensation of a water cloud may
increase the Bond albedo of a planet by up to a factor of two.
The spectra of cloudy planets are strongly influenced by poorly constrained
cloud microphysical properties, particularly particle size
and supersaturation.
Both Bond and geometric albedos are comparatively
less sensitive to variations in planet mass and effective temperature.

\end{abstract}


\section{Introduction}

The indirect detection of planets surrounding nearby
stars (Mayor \& Queloz 1995; Marcy  \& Butler 1996; Butler \& Marcy 1996; 
Butler et al. 1997; Cochran et al. 1997) has spurred efforts to both understand
the origin and evolution of these bodies and to directly
detect them.  Evolution models depend upon a computation of the
deposition of incident radiation into the planetary atmospheres
and calculations of planetary albedos.  Detection efforts require
estimates of the spectra of the extrasolar planets.  We
present here models of the reflected spectra and albedos
of extrasolar planets to facilitate both the modeling of
the evolution of these objects and ultimately their direct
detection.

While it is currently not known if the new objects are terrestrial planets,
jovian planets, or even brown dwarfs, we focus here on jovian planets
and brown dwarfs.   Our justification
is largely drawn from planetary formation models (e.g. Pollack et al.
1996), the cosmic abundance of the elements, and our prejudice from 
our  own solar system, supported by minimum masses
for the new objects exceeding $0.4\,\rm M_J$, where $\rm M_J$ is
one Jupiter mass.  Regardless of mass,
we refer to these objects as Extrasolar Giant Planets (EGPs).

Saumon et al. (1996) presented the first comprehensive study of the evolution 
and spectra of extrasolar giant planets.  They modeled the reflected component 
of the spectra as a graybody reflecting  incident blackbody radiation.   
As Saumon et al. recognized, this approach overestimates the reflected 
flux from Jupiter by up to two orders of magnitude in certain bands. 
Thus a complete, non-gray, model of atmospheric scattering is required
to fully model the reflected component of these atmospheres.

Burrows et al. (1997) reported on the evolution and emitted spectra of extrasolar jovian 
planets and brown dwarfs.  They modeled solar composition objects
and included the influence of neither incident sunlight nor clouds.  Their
thermal emission spectra extended from slightly shortwards of $1\,\rm \mu m$
to beyond $100\,\rm \mu m$.  To construct
boundary conditions for their evolution models and emergent
spectra, they computed radiative-convective equilibrium
atmosphere models for over 50 objects with effective
temperatures ranging from 128 to 1200 K and gravities between
22 and $3000 \,\rm m\,sec^{-2}$.  This phase space
covers the range from Jupiter to the most massive brown dwarfs.   
Burrows et al. found that the emergent flux
from such objects can deviate strikingly from a blackbody, with
fluxes in some bands orders of magnitude larger than those of a blackbody with the same effective
temperature.

Here we report on the reflected spectra and albedos of extrasolar planets and brown dwarfs.
To present as general a set of results as possible, we
calculate the geometric albedo spectrum of each model planet.
The actual reflected spectrum for a planet
around an arbitrary star can be computed from
the geometric albedo spectrum, the appropriate incident stellar spectrum, 
and the phase function of scatterers in the planetary atmosphere.  The geometric
albedo spectrum of Jupiter and Uranus in the optical are shown in Figure 1.

Condensates dramatically influence the reflected spectra
of all the substantial atmospheres of our solar system.  We explore the role of both silicate and water clouds in the reflected spectra of the extrasolar planets and brown
dwarfs and discuss potential tests for clouds.   However we postpone a
detailed investigation of the role of condensates to a future paper.

We stress that we are not attempting to exactly predict
the spectrum of a particular extrasolar planet.  No observed reflected
spectra yet exist for the extrasolar planets and their  masses, compositions, and
ages (and thus internal heat flows) are constrained poorly,
if at all.  Given  the large uncertainties in atmospheric composition, atmospheric
hazes, cloud decks, particle sizes and compositions,  not to mention
the poor quality of the high
temperature absorption coefficients of  many molecules of interest,  we choose to examine the overall phase
space of gravity and effective temperature within which the new planets
reside.  This work, in combination with Burrows et al. (1997), provides
the means to quickly estimate the reflected and emitted flux from an extrasolar
giant planet or brown dwarf.  
Detailed modeling of specific planets is retained for the future although
Guillot et al. (1997a) have presented some of our early results for specific planets.

\section {Albedo Computation}

\subsection {Theory}

The Bond or bolometric albedo $A$ is the ratio of the total power reflected
by a planet to the total power incident upon the planet,
$$A = {P_{\rm refl}\over P_{\rm incid}}.\eqno(1)$$
This albedo is used to compute the planetary energy balance: 
the effective temperature $T_{\rm eff}$ at which a planet radiates to space is
given by
$$4 \pi R^2 \sigma T^4_{\rm eff} = E_{\rm int} + \pi R^2(1-A)\pi S/d^2.\eqno(2)$$
Here $R$ is the radius of the planet, $d$ is the distance to the sun
in astronomical units, $\pi S$ is the solar constant or flux
at 1 AU, $\sigma$ the Stefan-Boltzmann constant, 
and $E_{\rm int}$ represents the internal energy sources
(e.g. Conrath et al. 1989).  The total reflected flux ${\cal F}_\nu$ received 
from an extrasolar giant planet  a distance $a$ from
Earth that orbits a star 
of radius $R_*$ is given by (e.g., Saumon et al. 1996)
$${\cal F}_\nu = {A\over4} P(\Phi) 
\Biggl({R_* \over d}\Biggr )^2 \Biggl( {R\over a}\Biggr )^2 {\cal F}_\nu^*.\eqno(3)$$
${\cal F}_\nu^*$ is the flux radiated by the surface of the star and $P(\Phi)$ is a function with value of order unity that accounts for the dependence of the reflected light upon phase angle $\Phi$, 
the angle between the star, the planet, and the observer.  

To compute the Bond albedo, expressions are required for 
$P_{\rm refl}$ and $P_{\rm incid}$.  The incident power is 
straightforward,
$$ P_{\rm incid}=\pi R^2 {{\pi S}\over{d^2}}.\eqno(4)$$
Neglecting any possible dependence of reflectivity on latitude
(e.g. polar caps), the total reflected flux can be computed from
the mean disk intensity $\bar I(\Phi)$ by an
integral over phase angle (Hanel et al. 1992),
$$P_{\rm refl}= 2 \pi^2 R^2 \int_0^\pi {\bar I}(\Phi) \sin \Phi d\Phi . \eqno(5)$$
After the division of Eq. (5) by Eq. (4) it is possible to write
$$A = pq = {{{\bar I}(0)}\over {S / d^2}}\cdot 2 \int_0^\pi 
{{{\bar I}(\Phi)}\over {{\bar I}(0)}} \sin \Phi d \Phi. \eqno(6)$$
This equation defines the second commonly used albedo, the
geometric albedo $p$ 
$$p={{\bar I}(0)\over {S/d^2}}.\eqno(7)$$
The geometric albedo is simply  the reflectivity of a planet
measured at opposition.  It can be expressed 
as either a mean value integrated over the solar spectrum, $\bar p$, or  as
a monochromatic value, $p_\lambda$.  
Spectra of Jovian planets are commonly presented as the latter.
For a perfectly-reflecting Lambert sphere
the geometric albedo is 2/3 (Hanel et al. 1992);  for a semi-infinite purely
Rayleigh scattering atmosphere it is 0.75 (Dlugach and Yanovitskij 1974).

The Bond albedo is then formally related to the monochromatic
geometric albedo (e.g. Pollack et al. 1986) by
$$A = \int_0^\infty p_\lambda q_\lambda f_\lambda d\lambda \Biggl/ \int_0^\infty
f_\lambda d\lambda \eqno(8) $$
where $q_\lambda$ is the 
monochromatic phase integral and $f_\lambda$ is the incident monochromatic
solar flux.  Following Eq. (6), the phase integral is defined by
$$q = 2  \int_0^\pi {{\bar I}(\Phi)\over {\bar I}(0)} \sin \Phi d\Phi.\eqno(9)$$
The phase integrals for a Lambert sphere and a Rayleigh atmosphere are  3/2 and 
4/3 respectively.

While $q_\lambda$ can vary with wavelength in a complicated
manner, it is common to remove an appropriately averaged
mean phase integral $\bar q$ from the integral and let
$$A = \bar q \int_0^\infty p_\lambda f_\lambda d\lambda \Biggl/ \int_0^\infty
f_\lambda d\lambda \equiv \bar q \bar p.\eqno(10)$$
With this approach the task of computing the Bond albedo falls
to separately computing the mean phase integral and mean geometric albedo.  
Pollack et al. (1986), for example, computed $\bar p$ and $\bar q$ for a wide range of possible
pre-Voyager Uranus atmosphere models, including models with strongly forward-scattering 
clouds and with isotropically-scattering clouds.  They found that $\bar q$
varied only by about $\pm 10\%$ among different models, with a typical value
of about 1.2.  This is because most reflected photons near the
peak of the solar Planck function are scattered by Rayleigh scattering,
and the phase integral for Rayleigh scattering with a moderate amount
of molecular absorption ranges between 1.25 and 1.30.

Unlike Pollack et al., after finding both $q_\lambda$ and $p_\lambda$ as a function
of wavelength  we compute $A$  via Eq. (8) rather than Eq. (10).  By varying
$f_\lambda$, we compute  Bond
albedos for a variety of stellar types along the main sequence in Section 4.2.

\subsection {Atmosphere Models}

To compute the  spectra for an atmosphere with a
given effective temperature and gravity requires a
model which describes an atmosphere with those
characteristics.  The most accurate way to proceed
would be to compute a self-consistent radiative-convective
equilibrium model atmosphere for each possible type of 
primary star, internal heat flow, orbital distance, and surface gravity.
Such a model would compute both the deposition of incident radiation 
with height and the transport of internal energy.  If the computed
temperature profile crossed condensation equilibria 
contours, clouds would be inserted and the deposition of
solar radiation recomputed.  Indeed this approach was
taken by Marley  and McKay (1998) to compute atmospheric
temperature profiles for Uranus and by Marley (1998) for exploratory
models of EGP atmospheres.

However, given the large phase space of models which
we wish to explore, we choose not to compute atmosphere
models which include the deposition of solar radiation.
Rather we use the temperature profiles presented in 
Burrows et al. (1997) which are computed for isolated
brown dwarfs and extrasolar planets.  In addition, several new, lower gravity models 
(down to $g= 5\,\rm m\,sec^{-2}$) were also computed for this work, 
following the techniques presented in Burrows et al. (1997). 

We then ask what these models would look like
under various illumination conditions.  In other words, we compute
the reflected spectra of ``isolated'' brown dwarfs and extrasolar planets.
In doing so, we neglect the effect of the incident radiation on
the temperature profile.  
Incident radiation would have two main effects on the models.  
Firstly, methane would absorb incident radiation above
the tropopause, producing a warm stratosphere as in the solar
Jovian atmospheres.  This behavior is seen in exploratory
extrasolar atmosphere models (Marley 1998).  
Except for those few objects where 
the tropopause temperature may rise sufficiently 
to prevent condensation, for most  objects  with a fixed effective 
temperature, the change in the temperature profile resulting from
absorption of the direct insolation should not  substantially alter 
the reflected spectra calculated here.  Secondly, incident UV
radiation will drive photochemical reactions that 
can produce stratospheric hazes.  As discussed in Section 3.4
such hazes will both absorb additional incident UV radiation
and scatter infrared photons thereby affecting both the
temperature profile and the reflected spectrum. Results from
a detailed investigation will be presented in a future paper.

Each atmosphere model consists of up to 53 levels, spaced
approximately evenly in the log of atmospheric pressure.
In each layer (layers describe the atmosphere between discrete levels) 
the single scattering albedo $\tilde \omega_0$  and
scattering asymmetry factor  are computed
for the relevant scatterers.  Mean values of these parameters
are  found for each layer by weighting each  constituent's scattering   parameters  by
its layer scattering optical depth.  Opacity sources are 
discussed in the next section.

The geometric albedo spectra are then computed following the approach
employed by McKay et al. (1989).  The source function from the Eddington 
two-stream approximation is used in an exact integral of the
radiative transfer equation.  This Eddington source function technique
and the associated errors are described in Toon et al. (1989).
The ratio of emergent to incident intensities are integrated over the disk
to compute the geometric albedo as a function of wavelength.
Previous applications of this code have reproduced the geometric albedo
spectrum of Titan and Uranus (McKay et al. 1989; Marley and McKay 1998).

For some models, particularly the warm, high gravity models, optical depth
unity at some wavelengths near $0.4$ to $0.5\,\rm \mu m$ is not reached
above the bottom of the model atmosphere.  This is because the column 
number density of absorbers and scatterers is 
low for such models. For these models we extend the
bottom of the model to 1,000 bars by extrapolating the temperature along
an adiabat.  Any remaining downward propagating photons at this level
are assumed lost.  

The phase integral is a function of the scattering properties
of the atmosphere.  For a conservative Rayleigh atmosphere $p = 1.33$.
As the single scattering albedo falls, 
multiple scattering becomes less significant, the geometric albedo decreases,
and $q$ decreases.  
Dlugach and Yanovitskij (1974) present tables
of $q_\lambda$ as a function of $\tilde \omega_0$ for purely Rayleigh
scattering atmospheres.  We compute the monochromatic
phase integral for the clear atmospheres by interpolation within these tables.
For a purely Rayleigh atmosphere the phase integral is not highly sensitive to
$\tilde \omega_0$. 
Over the tabulated range of $\tilde \omega_0$, $p_\lambda$ varies from 
0.75 to 0.2, while $q_\lambda$ varies only between 1.33 and 1.26.  For cases
of small $\tilde \omega_0$ that fall outside the tables, 
we set $q= 1.25$.  Since the low values of $\tilde \omega_0$ are found in the infrared where $f_\lambda$ is also  comparatively small, the exact
treatment of $q$ for small $p$ has  no influence on the Bond albedo.

Since the scattering phase function of cloudy planets differs from
a Rayleigh phase function, we compute $q$ for the cloudy cases by
interpolating within tables of Dlugach and Yanovitskij (1974) that
give $q$ as a function of both $\tilde \omega_0$ and the scattering
asymmetry parameter.  In these cases the scattering behavior depends on whether or not the clouds are visible at a given wavelength. For wavelengths in which the normal scattering optical depth reaches unity  above the cloud layer, we proceed as above for a purely Rayleigh
atmosphere since the clouds are not visible.  
When most photons are scattered in the cloud, we compute the mean
layer asymmetry factor and single scattering albedo at that wavelength for the layer in which the normal scattering optical depth reaches unity.  We then interpolate to find $q_\lambda$.  Some sample $q_\lambda$ spectra
are shown in Figure 2.  Since $q_\lambda$ varies only weakly between models,
we do not present multiple plots of this parameter.

Our approach for a purely Rayleigh scattering, absorbing  atmosphere is near exact.  For the
three cloudy atmospheres we consider here, the computed values of $q_\lambda$ are estimated to be accurate within 10\%.  In a future paper we will consider a broader range of cloud models and present an exact calculation of $q_\lambda$.

\section {Atmospheric Opacity Sources}

The atmosphere  models presented in Burrows et al. (1997) assumed
solar elemental composition.  Chemical equilibrium abundances
for all species are computed for each model layer  and a
self-consistent radiative-convective equilibrium temperature
profile is derived.  For these same models 
we compute the reflected radiation.  In this section we summarize
the major contributors to atmospheric opacity at wavelengths
shorter than $5\,\rm \mu m$.

\subsection {Rayleigh Scattering}

For a semi-infinite, purely molecular atmosphere, all photons received by an
observer have been returned by Rayleigh scattering.  A
rigorous treatment of Rayleigh scattering would employ the
full phase matrices to compute the intensity of scattered
light.  Indeed Pollack et al. (1986) investigated the
difference in computed geometric albedos for calculations
which employed the full phase matrix and those which simply
used a scalar phase function.   They found that in
an infinitely-deep purely Rayleigh atmosphere the scalar phase
function treatment of Rayleigh scattering resulted in an
under-estimate of the geometric albedo by about 6\%.

Pollack et al. employed an empirical correction scheme to
account for this discrepancy.  We choose instead to normalize
our treatment of the Rayleigh phase  integral so  that our
code returns the correct geometric albedo (0.75) in the limit of 
an infinitely-deep purely Rayleigh atmosphere.

\subsection{Raman Scattering}

Belton et al. (1971) first recognized the importance of
Raman scattering in decreasing the ultraviolet geometric albedo
in deep, Rayleigh scattering planetary atmospheres.  During Rayleigh
scattering a fraction of photons
excite vibrational and rotational transitions of 
$\rm H_2$.  The scattered photons thus experience a shift
to longer wavelengths.  Excitation of rotational transitions
produces very small wavelength shifts and is responsible
for ``Raman ghosts'' of solar Fraunhoffer lines.  We
are concerned instead with the much larger shifts arising
from excitation of vibrational transitions, which removes
photons from the UV and blue.

Pollack et al. (1986) discuss several approximate and exact
techniques for computing the effect of Raman scattering on
the geometric albedo spectrum.  They
find that in a deep $\rm H_2$ atmosphere Raman scattering can decrease
the geometric albedo by 20\% at $0.3\,\rm \mu m$.  We choose
the same approximation for Raman scattering as adopted by
Pollack et al.  and express the mean single scattering albedo $\tilde \omega_0$
of a layer as
$$\tilde\omega_0 = {{\sigma_{\rm R} + \sigma^\prime_{\rm S} + \sigma_{\rm RA}
(f_{\lambda^*}/f_\lambda)}
\over {\sigma_{\rm R} + \sigma^\prime_{\rm e} + \sigma_{\rm RA}}}.\eqno(11)$$
Here $\sigma$ denotes the cross sections for
Rayleigh scattering (subscript R), Raman scattering (RA), scatterers other than
$\rm H_2$ (S), and other sources of extinction (e).  The ratio
$f_{\lambda^*}/f_\lambda$ accounts for the wavelength dependence
of the incident solar spectra. The wavelength of interest is
denoted by $\lambda$ and $\lambda^{*(-1)} = \lambda^{-1} + \Delta\lambda^{-1}$,
where $\Delta \lambda$ is the wavelength of the $\rm H_2$ 
vibrational fundamental $\Delta \lambda ^{-1} = 4161\,\rm cm^{-1}$.
The Raman correction is confined to wavelengths less than the
peak of the Planck function, thus limiting the maximum single
scattering albedo to be less than 1.
Pollack et al. (1986) find that geometric albedos computed with
this approximation match well those computed with an exact treatment
of Raman scattering.

The geometric albedo spectra presented in Section 4 are computed
using values of $f_{\lambda^*}/f_\lambda$ computed for a 6000 K blackbody,
an adequate approximation since we are most interested in the
general shape of the spectrum.
For a rigorous calculation of the spectra of a particular
planet, the flux ratio appropriate to the particular primary must be used.
Over the range of spectral types considered here, this approximation changes $p_\lambda$ by only a few 
percent at the shortest wavelengths.  For very early type stars (O and B)
the stellar Planck function peaks in the ultraviolet and Raman scattering
would have the opposite sign in the UV, but such models are not considered here.

\subsection{Molecular Opacities}

The sources of molecular opacity data and the treatment of line broadening is discussed fully in Burrows et al. (1997).  Our approach differs only in that we do not treat 
opacities with the  k-coefficient technique, but rather compute exact fluxes on a fixed wavelength grid.  The opacity within a given atmospheric layer  is a function of its composition, temperature, and pressure.  We interpolate within opacity tables computed for  fixed conditions within the model to find the opacity of a given species at arbitrary temperature and pressure.

In the spectral region considered here, the most important  molecular opacities are those of water, ammonia, methane, and pressure-induced  absorption by hydrogen.  Ammonia bands lie near  0.64 and $0.78\,\rm\mu m$ and  between 1.4 and $1.55\,\rm\mu m$,  1.9 and $2.05\,\rm\mu m$, and  2.2 and $2.4\,\rm\mu m$.
The major methane absorption bands lie at 0.725, 0.89, 1.0, 1.15, 1.4, 1.7, and $2.5\,\rm\mu m$ with many other weaker bands in the optical.  Water
bands  are important for the warmer objects considered here.  The strongest bands in the optical lie at 0.83 and $0.95\,\rm\mu m$.  Stronger bands fall at  1.1, 1.4, 1.9, and $2.9\,\rm\mu m$.  The fundamental of the pressure-induced vibrational band of  molecular hydrogen is centered at $2.4\,\rm\mu m$ and the first overtone is at  $1.2\,\rm\mu m$.  Several other weaker  hydrogen features populate the visible.

\subsection{ Photochemical Hazes}

As is apparent for Jupiter and Uranus in Figure 1, the geometric albedo 
in the UV and blue for all the solar Jovian
planets is lower than predicted for  purely Rayleigh (and Raman)
scattering atmospheres (e.g. Savage and Caldwell 1974).  
Hazes created by the photochemical destruction
of methane and other molecules produce a smog of particles that
are dark in the blue and ultraviolet.  There is a rich literature
of observations and models of such hazes in planetary atmospheres.  Indeed
these hazes were first recognized from the depressed UV and blue
geometric albedos of the planets.  Furthermore the photochemical
products themselves (e.g. $\rm C_2H_2$ and $\rm C_2H_6$) also
absorb in the ultraviolet.

While the presence of the hazes is known and their origin
well understood, they are particularly difficult to model.  Their
size and vertical distribution depends on photochemical reaction
rates, eddy mixing coefficients, and nucleation and
coagulation rates that are poorly known.  In warm extrasolar atmospheres rich in 
 C, N, O, and S bearing molecules, it is likely that photochemistry will
yield a particulary  complex  brew of non-equilibrium compounds
(Marley 1998).  Both the compounds themselves and any resulting
condensates may lower the reflected UV flux.  Indeed even recent attempts
to model the hazes detected by Voyager have met with only partial
success (e.g. Rages et al. 1991, Moses et al. 1995).  Thus we
neglect photochemistry in our model atmospheres
while recognizing that its influence may be significant.  Photochemical hazes
hazes would likely depress the geometric albedo in the blue and UV
while increasing the albedo in the near-infrared molecular bands
of water and methane (by scattering photons before they are absorbed).
As a result the Bond albedos for planets orbiting early type stars
will be quite sensitive to the effects of photochemistry.

\subsection{Clouds}

We also consider the role of clouds in affecting planetary
reflected spectra.
A cursory look at any of the planets in our solar system 
with a substantial atmosphere, 
including Earth, reveals that clouds  dramatically affect the visible appearance of planets.   
Guillot et al. (1997b) have explored condensation processes in the atmosphere of 
extrasolar giant planets.  They find that at effective temperatures above 
about 1100K silicates, including $\rm MgSiO_3$, condense in the atmosphere.  At 
effective temperatures below about  400 K, water condenses.  At intermediate 
temperatures lower abundance species, such as $\rm Na_2S$ and ZnS condense.  
At the very cool temperatures characterizing the atmospheres of our own 
jovian planets $\rm NH_4SH$, $\rm NH_3$, and $\rm CH_4$ also condense.  
Thus all of the atmospheres considered in this paper likely  contain hazes or clouds of various species.  There
is no doubt that these condensates will alter the reflected spectra of extrasolar giant planets.
To explore the potential effects of these condensates we will concentrate on the most abundant condensates, water and enstatite, $\rm MgSiO_3$.  

In this section we discuss models for the physical size and vertical
extent of water and silicate clouds and consider their effect
on the atmospheric opacity.

\subsubsection{Cloud Profiles}

The characterization of the physical properties and radiative
effects of condensed species in an atmosphere is a difficult
exercise unless extensive observations are available. Experience
with the Earth shows that cloud particle sizes, vertical and
horizontal distributions are sensitive functions of the ambient
atmospheric conditions. In the terrestrial atmosphere a strong feedback exists between
cloud evolution and the background temperature, winds and
vertical turbulent motions. On the other hand, the simplest
model in which the vapor pressure relationship of the condensed
species is used to predict cloud properties is inadequate because
no information on condensate particle size is available from this
equation by itself. 

Ultimately detailed fitting of high-resolution spectra will
enable extraction of information on the globally-averaged cloud
properties of extrasolar planets and brown dwarfs. As a start to
this process, we have constructed a simple scheme to predict the
particle size as a function of altitude for two condensible
species, water and magnesium silicate, in two end-member
environments corresponding to quiescent and turbulent atmospheric
states. The model builds on an earlier scheme  in Lunine et al.
(1989). We begin with the cloud-free atmospheric models whose
construction is described above, and use the temperature-pressure
profile (along with assumption of roughly local solar elemental
abundance) to determine the cloud base and vapor abundance as a
function of altitude. Vapor pressure relations for water (liquid
and solid) are from Eisenberg and Kauzmann (1969), for solid
magnesium silicate from Barshay and Lewis (1976). 

Given the vapor pressure abundance we compute
a maximum mixing ratio of condensate in the cloud-forming region
by taking the vapor abundance and multiplying by a suitable
supersaturation factor, which yields the amount of condensible
available for condensation. 
The supersaturation is formally defined as $f_s=(P_v/P_s)-1$,
where $P_v$ is the actual vapor pressure reached prior to condensation
and $P_s$ is
the thermodynamic saturation vapor pressure of water or silicates over
the same type of condensate (ie., solid or liquid).  
The supersaturation varies widely
among and within planetary atmospheres, and is a sensitive
function of the local properties of condensation nuclei which are
not well known for extrasolar planets; generally the
supersaturation for a given condensable  species is larger in
colder atmospheres and in the relative absence of nucleating
aerosols. As a baseline we choose $f_s=0.01$,
so that the condensate abundance at each atmospheric altitude is
1\% the vapor phase abundance. This number is simply a typical
value which fits some terrestrial situations involving
condensation onto nucleating sites, but needs to be recomputed in
actual fitting of high resolution observational spectra to
results of spectral synthesis models (see, for example the
Courtin et al. (1995) analysis for Titan). For the given
supersaturation factor we regard our condensate abundances as
practical upper limits because physical processes such as sedimentation,
atmospheric downwelling, etc., will tend to decrease the
globally-averaged column abundance of cloud particles.  However
to evaluate the sensitivity of the computed spectra to this
parameter, we also consider cases with other values of $f_s$.

To estimate the cloud particle size we follow one of two approaches.
In the first, we employ a simple model to compute the variation
of cloud particle size with height in quiescent and turbulent
atmospheres.  In the second we explore the spectral sensitivity
to changes in particle size
by considering a suite of particle sizes independent of any 
condensation model.

For the first approach, we employ the formalism of Rossow
(1978) who computed timescales associated with grain growth by
condensation and agglomeration, and loss by sedimentation. The
equations for the various growth and loss processes are given in
Rossow (1978) and will not be reproduced here. We further
consider two endmember atmospheric situations.  A {\it quiescent
} atmosphere is one in which turbulence does not generate
macroscopic eddy motions which serve to keep relatively large
particles from rapidly sedimenting out of the cloud layer. In the
quiescent atmosphere we balance, as a function of particle size,
the growth rate (by condensation or coagulation, whichever is
faster), against the sedimentation rate. Particles large enough
such that the sedimentation rate just exceeds the growth rate are
assumed to be the modal particle size. Since we do this at each
level within the cloud we get a profile of cloud particle size
and cloud particle density throughout the cloud. For the {\it
turbulent} atmosphere the same procedure is followed, but here we
compute the rate of eddy mixing and balance it against the
sedimentation rate as a function of particle size. Particles
large enough that the sedimentation rate just exceeds the rate
of remixing by eddy turbulence are again assumed to fall out of
the cloud. The eddy mixing rate is 
$\bigl[({\cal R}\sigma T_{\rm eff}^4)/(\rho {\bar \mu} c_p)\bigr]^{1/3}(3H)^{-1}$, 
where $H$ is scale height, $\cal R$ the gas constant,
$\bar \mu$ the atmosphere's mean molecular weight, $\rho$
its density and $c_p$ its specific heat. 

The radiative-convective atmosphere models   describe
which regions of the atmosphere are convective
and which are radiative, thus presumably predicting whether
the turbulent or quiescent cloud model is most applicable to
a given situation.
However, these calculations are made for a clear atmosphere.
The presence of the cloud layers themselves will change the predicted
profile, so we consider grain growth in both environments.

The resulting particle sizes as a function of altitude for the turbulent
case are significantly larger than for the quiescent atmosphere, in
agreement with terrestrial observations.  Figures 3a and b 
illustrate the physical
properties of the cloud models.  For a typical model 
designed to approximate  the brown dwarf $\rm G\ell\,229\,B$, with effective temperature of 1000 K and roughly 40 Jupiter masses, enstatite
(MgSiO$_3$) clouds
have their base at 48 bars. In the quiescent case maximum
particle sizes near the base are 30 microns at the base dropping
to 2 microns higher in the cloud. In the turbulent case particle
radii have a very narrow maximum size range throughout the cloud
of between 100 and 300 microns.  

To better elucidate the dependence of the computed spectra
on particle size, we also explore clouds of various uniform,
fixed, sizes.  Atmospheric condensates found in the solar system
range from submicron photochemical 
hazes to micron-sized methane clouds (Rages et al. 1991),
to even larger drops.  Indeed the upper particle 
size limit for even Jupiter's clouds is not well constrained.
We thus consider clouds with fixed radii varying from 0.1 to $100\,\rm \mu m$.

\subsubsection{Mie Scattering}

The radiative effects of clouds were modeled by Mie scattering
theory.  The extinction and scattering efficiencies, single
scattering albedos, and scattering asymmetry parameters were computed for
spherical drops of water and enstatite.  Mie scattering computations for drops of a single particle size show numerous, very high frequency, spectral features resulting from constructive and destructive interference of  radiation inside 
and around an idealized drop.  Since real
clouds are composed of drops of a range of sizes, such fine structure  becomes washed out
and is seldom observed.  We thus assume that the model clouds are characterized
by a log-normal size distribution with a width parameter of 1.5.  
This size distribution is motivated by observations of real clouds 
in a variety of planetary atmospheres (e.g. Venus, Titan, and Uranus) 
and was chosen because the high frequency  spectral features 
are removed while preserving the most physically significant variations 
in particle scattering parameters with wavelength. Detailed models 
of cloud particle nucleation, evaporation, sedimentation, 
and coagulation can motivate  particular choices for the width parameter, 
but until such studies are motivated by exceptionally  
detailed spectra of extrasolar planets, such concerns are as 
yet premature. Other choices for the particle size distribution
function are likewise possible, but all choices are currently equally
unmotivated at this time by data or theory.

More complex treatments of scattering by condensates are possible.
For example rhomboid or even fractal particles can be
considered.  The influence of such non-spherical
particles primarily is apparent in the scattering asymmetry.  
However such treatments are usually only justified when the radiative
transfer problem is exceptionally well constrained, generally
uncertainties in Mie scattering are far less than the other
unconstrained aspects of the problem.  
As the cloud particle
sizes and shapes are essentially unconstrained, going beyond
Mie theory is unwarranted.

The choice of composition and particle size determines the Mie
scattering parameters.  For water we use the
optical properties of Hudgins et al. (1993).  
For the silicate clouds we employed the optical properties of
amorphous enstatite from Scott and Duley (1996).
The extinction optical depth of a model
layer is then given by 
$$\tau = \pi  r_c^2  Q_{\rm ext} \cal N \eqno (10)$$
where $r_c$ is the mean layer cloud particle size, $\cal N$
is the layer column number density, and $Q_{\rm ext}$ is
the Mie extinction efficiency for the given particle composition
and size.  Scattering and
absorption optical depths are similarly computed.
Mie theory predicts that the extinction efficiency at optical wavelengths
will be essentially constant for all particles with $r_c > 1\,\rm \mu m$.
Since the total condensible mass
in the atmosphere is fixed for a given model, ${\cal N} \propto r_c^{-3}$ and
$\tau \propto r_c^{-1}$ for $r_c> 1\,\rm \mu m$.  

\subsection{Stellar Spectra}

Fourteen stellar spectra (spectral type A5V-M6V) were taken from the 
Bruzual-Persson-Gunn-Stryker Spectrophotometry Atlas (Space Telescope Science
Institute Data Analysis System; Gunn and Stryker 1983; Strecker et al. 1979).  
Spectra from the atlas are normalized to V magnitude 0.  We renormalized
each spectra by multiplying the spectral flux $f_\lambda$ by the ratio 
of $\sigma T_{\rm eff,*}^4$
to the integrated atlas flux, where $T_{\rm eff,*}$ is the stellar
effective temperature.  Stellar fluxes were extended beyond $2.56\,\rm \mu m$,
the limit of the atlas, with a Planck function of the
same effective temperature.  The integrated albedos are not
sensitive to the details of the flux at these wavelengths.

Reflected fluxes are computed for a planet 1 AU from its primary star
with radius from the table in Lang (1992).  Planetary radii for
each model were derived from scaling relations in Marley et al. (1996).  Spectra
are computed by multiplying the incident stellar flux by the
geometric albedo. For objects  observed at arbitrary phase,
these spectra must be modified to account for the partial illumination
of the disk and the phase function of atmospheric scatterers.  For
an object at half phase ($\Phi=90^\circ$) the total correction factor would be
near 0.6.  

\section{Model Results}

\subsection{Geometric Albedo Spectra}

\subsubsection{Planetary Spectra}

Before discussing model results, we first compare the predictions
of our model to the known Jovian planet spectra.  
Figure 1 displays the geometric albedo spectrum of both a cloud-free
and a cloudy model
with Jupiter's gravity and effective temperature.
Also shown are the geometric albedo spectra of Jupiter and Uranus.  
All these spectra are dominated by the optical bands
of methane.  However the clear model spectrum appears more similar
to the spectrum of Uranus than that of Jupiter.  This is because
Uranus' visible atmosphere is clearer, deeper, and
contains 10 times more methane (Lindal 1987) than Jupiter's.
These characteristics, coupled with the planet's lower gravity, result in a 
40-fold larger column abundance of $\rm CH_4$ above the uppermost
cloud deck than in Jupiter's atmosphere.  Since the clear model
does not have a lower reflecting boundary, the model appears more
similar to Uranus than to Jupiter.  The cloudy model, with
its high aerosol haze and thick cloud is meant to only roughly
represent Jupiter.  An exact fit requires the fine tuning
of far more model parameters than we explore here.

Scattering from bright clouds also accounts for the brightness
of both Jupiter and Uranus in-between methane absorption bands
beyond $0.6\,\rm \mu m$.
In Jupiter's atmosphere upper cloud decks of $\rm NH_3$
and $\rm NH_4SH$ scatter incident radiation back to space before it
can be absorbed.  In the perfectly clear model atmosphere downward photons
in the red and beyond are lost to weak molecular absorption long before 
they can Rayleigh
scatter back to space.  Thus the clear model is almost black beyond
$1\,\rm \mu m$ while Jupiter, Uranus, and the cloudy model remain
bright between the strong
$\rm CH_4$ and $\rm H_2 - H_2$ bands which define the near-infrared
spectra. Indeed, were it not for its clouds and hazes, Jupiter would be essentially undetectable  in reflected light in the near infrared.  

At wavelengths less than about  $0.65\,\rm\mu m$ Jupiter is darker
than the model.  As discussed in Section 3.4, this difference is 
attributable to the presence of absorbing
hazes high in Jupiter's stratosphere.  These hazes lower the mean single
scattering albedo of the atmosphere below that expected for  a purely
Rayleigh and Raman scattering atmosphere.   Some thin UV-dark hazes also
somewhat depress the UV albedo in
Uranus' atmosphere.  Again, however, the clear, Rayleigh
and Raman scattering model atmosphere is more similar to Uranus
than to Jupiter.  Jupiter's albedo is also lowered in this
wavelength region by absorption by other hydrocarbons. The bright
cloudy model scatters some UV photons before they can be Raman
scattered and is thus bright in the UV.

Finally, many of the high frequency features seen
in both Jupiter's and Uranus' spectra are Raman ghosts,
discussed in Section 3.2.  These features do not appear
in the model spectrum since we neglect Raman rotational scattering.

As the above comparisons suggest, only models optimized to fit
an individual planet's atmosphere
can hope to exactly match the
spectra of a given planet. The planetary
spectra in Figure 1 can be reproduced exactly only by including
multiple cloud and haze layers with wavelength dependent
scattering properties (e.g. Baines and Bergstralh 1986).
These quantities are determined 
by carefully fitting the observed spectra.  There is no
theory that {\it predicts} these quantities.  Given these
empirical difficulties, our approach is simply to explore
the sensitivity of planetary albedo to various model parameters.

\subsubsection{Cloud-free EGP Spectra}

Model geometric albedo spectra for clear and cloudy atmospheres are shown
in Figures 4 through 7.  In each case the spectra were computed
for 3,300 points between 0.3 and $5\,\rm \mu m$ and then
smoothed with a Gaussian to produce the figures.

The sensitivity of the cloud-free model spectra to gravity
and effective temperature is explored in Figures 4 and 5.
Figure 4 compares the spectra of three models with Jupiter's effective 
temperature but differing gravity.   These spectra all demonstrate 
that deep, cloud-free atmospheres are remarkably dark in reflected
light beyond about $0.6\,\rm \mu m$.
Among the models is very little sensitivity to 
changes in gravity.  The main trend is that the reflectivity falls
with increasing gravity.  This and the other differences arise from 
the differing temperature 
profiles (different number densities at a fixed pressure), and relative 
differences in the strength of molecular absorption and Rayleigh scattering.

Figure 5 compares the geometric albedo of two cloud-free  models with Jupiter's gravity,
but differing $T_{\rm eff}$.  These correspond to objects with
masses of about  2 and 3 $\rm M_J$.  Again there  are
relatively few differences in the two spectra.  The principle differences are the appearance of an ammonia band at $0.62\,\rm \mu m$ and some differences
in band depths.  In the warmer atmosphere nitrogen is 
present as $\rm N_2$ and a pressure-induced  $\rm H_2$ absorption feature
near $0.82\,\rm \mu m$
is more prominent in the cooler, denser model since the absorption is proportional to the square of the number density.  The presence of
ammonia thus serves as a temperature discriminant.  As temperatures
drop further water condenses and the
water bands becomes less prominent in the optical.  In the
near-infrared, however, where photons penetrate more deeply, water
continues to be an important absorber.

As these results demonstrate, the cloud-free EGP spectra are relatively
insensitive to changes in gravity and effective temperature.  The
main spectral indicators for objects cooler than about $1200\,\rm K$
are first the appearance of the relatively subtle $\rm NH_3$ bands,
which start to appear for $T_{\rm eff} < 1000\,\rm K$ and the
disappearance of the water bands below about 400 K.  As discussed
in the next section, clouds leave a far greater imprint on the spectra.

\subsubsection{Cloudy Spectra}

The spectra presented in the previous section are limiting
cases since real atmospheres will certainly have condensates.
Guillot et al. (1997b) and Burrows and Sharp (1998) have investigated 
condensation in EGP atmospheres. They find that
water condenses in the atmospheres  of objects  with $T_{\rm eff}< 400\,\rm K$.
Although there will be some condensates  present in warmer atmospheres 
(section 4.1), the sudden appearance of bright water clouds will 
dramatically alter the reflected spectra of a planet cooling through 
400 K.   Silicate clouds will form in the observable portion of the
atmosphere only for objects with
effective temperatures above about 1000 K

The influence of water clouds is illustrated in Figures 6a, b, and c.
The quiescent and turbulent (Fig 6a) spectra correspond
to the cloud models shown in Figure 3a for a $2\rm M_J$, $300\,\rm K$
planet.  In Figure 6a the atmosphere is assumed to be highly supersaturated
($f_s=1$) to demonstrate the maximum influence of clouds.  
In this model the water clouds produce
a dramatic increase in the reflected flux beyond about $0.6\,\rm \mu m$.
by providing a bright reflecting layer high the atmosphere that scatters 
red and near-infrared photons before they have a chance to be absorbed.  
Longwards of about $1.5\,\rm \mu m$ molecular absorption above the cloud
tops is sufficiently strong to darken the planet despite the presence 
of the cloud.  Since the cloud particle size increases with depth into the 
cloud (from 2 to $128\,\rm \mu m$  at the cloud base for the radiative cloud), 
photons will scatter from the cloud at progressively deeper levels with 
increasing wavelength,  providing greater opportunity for absorption 
before scattering.  The difference between the two cloud models is a 
product both of the differing scattering properties of the differing particle 
size and, primarily, 
the differing particle number density attributable to the differing 
drop volume     in the two models.

The importance of the supersaturation factor is demonstrated by
Figure 6b.  For each decade drop in $f_s$ the cloud particle
column number density and the cloud optical depth (Eq. 10) 
proportionately fall.  Beyond $1.5\,\rm \mu m$ the residual
cloudy albedo is still many orders of magnitude larger
than  that of the cloud free (Figure 4) atmosphere.  
In the optical the albedo is clearly sensitive to 
the supersaturation factor.  For this case the $f_s = 0.1$
and 0.01 models are almost identical to the cloud free model.
This is not a general result as the total cloud column
optical depth will vary with effective
temperature, gravity, and cloud model.  Nevertheless the
strong sensitivity to the unknown $f_s$ is unmistakable.

In Figure 6c the quiescent and turbulent clouds of Figure 3a
have been replaced with various single particle sizes.  
Several factors are at work in the strong dependence of the
spectra upon cloud particle size evident in this plot.
For the smaller particle sizes there are many more drops,
thus increasing the cloud optical depth and the albedo (see
Section 3.5.2).  The $1\,\rm \mu m$-drops are most
abundant of all and also are far more efficient Mie scatterers.
Thus, like $f_s$  the cloud particle size plays a decisive
role in the planetary albedo.

Taken together, Figures 6a, b, and c demonstrate that any prediction of
the importance of clouds in an EGP atmosphere must account
for cloud microphysics, particle size, and supersaturation.  Since
all these quantities are completely unconstrained, only
ranges of possible spectra can be predicted for an arbitrary
planet.  

The effects of silicate clouds on the reflected spectra of a brown dwarf are shown in Figures 7a, b, and c.  
In this case the silicates darken the brown dwarf in the visible and
are less detectable in the infrared.  Compared to the water cloud, the
silicate cloud lies much deeper in the atmosphere (Figures 3a and b)
and infrared photons do not penetrate deeply enough to sense
the cloud.  However in the visible some photons do multiply 
scatter to the depth of the
the cloud deck.  Since the silicate 
grains do not scatter conservatively, the geometric albedo is lower for the
cloudy cases.  The visibility of the silicate cloud also demonstrates the smaller
scale height of the model atmosphere for the more  massive object.

The large gravities of massive extrasolar giant planets and brown dwarfs often defy the intuition gained 
from studying the solar Jovian planets.  Consider a cool brown dwarf similar 
to Jupiter, but more massive.  
It will have almost the same radius and thus a larger gravity.   The large gravity means that to compress  
a parcel of the atmosphere to a given pressure, a smaller column number density of molecules is required. 
In other words, given identical pressure-temperature   relations, the scale height  for the atmosphere of the more massive planet is everywhere smaller.  Thus given the same  atmospheric composition,  unity optical depth is reached at a 
larger pressure in the more massive  planet's atmosphere.  This is 
why it is possible to see the effects of a cloud which lies at 40 bars in a $36\,\rm M_J$ object, but not in Jupiter's atmosphere.

The effect of $f_s$ on the  quiescent cloud is demonstrated in 
Figure 7b.  As with the water cloud the influence of the cloud
depends sensitively on the total cloud optical depth and
consequently $f_s$.  The sensitivity to particle size is
presented in Figure 7c.  The submicron cloud both absorbs more
efficiently at short wavelengths and scatters more efficiently
in the infrared.

As these results demonstrate, in the
near-infrared any scatterer, even ones with low $\tilde\omega_0$
can substantially brighten the planet by reflecting incident light
before it can be absorbed.  The dark hydrocarbon hazes in the
atmospheres of all four solar Jovian planets are well known
examples.  The impact scars left by the fragments of comet
Shoemaker-Levy/9, which were dark at visible wavelengths and
extraordinarily bright in the near-infrared, are other examples.  
 The depth and shape of near-infrared absorption bands will thus
provide a powerful constraint on the nature of clouds and aerosols
in the extrasolar atmospheres.

\subsection{Bond  Albedos}

Table I lists Bond albedos computed for a variety of cloud-free 
extrasolar giant planet  
model atmospheres and primary types.  The masses of each object  with a 
specified effective temperature and gravity are estimated using the approximate
fitting relation given in Marley et al. (1996).  This expression
does not fit the lowest mass objects precisely (it predicts $M=2\,\rm M_J$
for Jupiter's gravity and $T_{\rm eff}$), and so the masses 
should be viewed only as a guide.  Bond albedos for the cloudy
models are shown in Table 2.  This table presents albedos for
both $f_s=0.01$ and 0.1 as well as for all the cloud types
discussed above.

Consistent with the spectra presented in Figures 4 and 5,
the Bond albedos in Table 1 show very little sensitivity
to gravity or effective temperature.  For a given stellar
primary the Bond albedo is constant to within about 10\%
for objects that vary by an order of magnitude in temperature
and a factor of 100 in mass.  This variation is comparable to
or even smaller than the uncertainty in the Bond albedos for
the solar Jovian planets (Conrath et al. 1989).  The poorly
constrained phase integral is most responsible for
the large error bars for the Jovian planets.

The large range in possible cloud models translates to
a very large range in Bond albedos in Table 2.  Depending
on the model, clouds can increase the Bond albedo by as much as a factor
of 2.   These variations are more clearly presented in Figures 8 and 9.

These figures show the possible variation in Bond albedo 
for  Jupiter mass (Figure 8) and brown dwarf mass (Figure 9)
objects.  The general trend 
for all models is that the later the primary type, the lower the Bond albedo.  
The  origin of this trend is immediately apparent from the geometric 
albedo spectra shown 
in the preceding section.  Because of the decrease in the strength 
of Rayleigh scattering  and the increase in the strength of molecular  
opacities, particularly methane and water, with wavelength, extrasolar 
Jovian planets are darker in reflected light in the infrared than they are 
in the visible.  The Bond albedo 
is defined, via Eq.~(8) as a weighted average of $p_\lambda q_\lambda$ over 
the incident flux.   Since the Planck function of later type stars peaks at 
progressively longer wavelengths, the Bond albedo falls with the stellar 
type of the primary  as one goes to later types along the main sequence.

These figures also demonstrate the great influence of clouds.
The Bond albedo can vary by a factor of two for a fixed primary type when clouds are added to the 
atmosphere.   As with the geometric albedo, the large sensitivity to cloud type 
arises from a combination of differing scattering properties and different column number densities 
for the various cloud models.

\section{Discussion}

The spectra of extrasolar giant planets have commonly been estimated by assuming they consist of two 
components:  a blackbody spectrum of the primary 
reflected by a gray surface and a blackbody thermal emission.   The spectra of all 
of the Jovian planets in our own solar system deviate widely from such a simple model.  
Thus it comes as no surprise that the model spectra for extrasolar planets also 
depart from such a simple picture.  

In reflected light  the planets do not resemble gray reflecting spheres.  
Instead their reflected spectrum is controlled by Rayleigh scattering
and molecular absorption.  In addition in the UV
absorption by hazes and non-equilibrium photochemically-produced
molecules will depresses the
reflected flux.  For all but the earliest spectral types (O and B)
Raman scattering also lowers the albedo for wavelengths less
than that of the peak stellar emission.  As a consequence
planets reflect most efficiently shortward of about $0.6\,\rm \mu m$
where photons
can Rayleigh scatter before they are absorbed.  At progressively longer
wavelengths extrasolar giant planets become darker as Rayleigh scattering
gives way to molecular absorption.  

In the red and near-infrared planets
are bright in reflected light only if there are clouds.  
Planets warmer than about 400 K will be dark
in reflected light in the red and near-infrared and have
relatively low Bond albedos.  As the planet cools through about 
400K, water clouds will appear, the planets will brighten, and
the Bond albedo will increase, perhaps dramatically.  The larger
Bond albedo will then hasten the cooling of the planet as less
incident stellar energy is absorbed.  Some condensates will likely always
be present in the warmer atmospheres (Guillot et al. 1997b). However all
such condensates are of species substantially less abundant than water
and will have a proportionately smaller influence on the reflected spectrum.

The great sensitivity of the geometric and Bond albedos to unconstrained
model parameters means that it will only be possible to
predict families of possible spectra for a given cloudy planet. 
The uncertainty is less for objects too warm to have condensed clouds
or for cases where the cloud lies relatively deep in the atmosphere.  The
great range of possible cloud models shown in Figures 6 and 8  is
partly a consequence of the high altitude of the water cloud in these
particular cases.  This same sensitivity, however, 
will allow the construction of detailed atmosphere models once spectra
are obtained of extrasolar giant planets.  It will then be possible
to study the physics of cloud formation in a large range of atmospheres.

The geometric albedo and phase integral spectra produced by this 
study can be combined with the 
emergent spectra of any primary star to generate an approximate
reflected spectrum.  Figures 10 and 11 demonstrate such a calculation
for objects orbiting at 1 AU from a G2V primary.  Again
adding water clouds to the cooler produces a substantial brightening.
The effect of silicate clouds is more subtle and is shown in Figure 11.

Burrows et al. (1997) demonstrated 
the emitted spectra of extrasolar giant planets are remarkably non-Planckian.  The planets 
emit  strongly in the windows in between strong methane and water absorption bands.  
Jupiter's own five-micron window is an example of this process.
Thermal emission spectra from the cloud-free models of Burrows et al. are also plotted
on Figures 10 and 11.  For the cloudy model, the thermal emission in the near-infrared
model is comparable to the reflected flux.  
For even warmer objects, the thermal emission will dominate the reflected
flux for objects at 1 AU and farther from their primaries.  For cooler objects 
the reflection from bright clouds will 
swamp the rapidly falling near-infrared emission.  For such
objects a full radiative transfer treatment of the combined
scattered and emitted radiation is clearly required.

The dominance of thermal emission for a hot planet at 1 AU is apparent 
in Figure 11.  The reflected flux would be comparable to the thermal emission only
for objects closer than about 0.1 AU to their primary.  Hence
the spectra of the 51 Peg B class of planets will be a mixture of
reflected and emitted radiation.  The relative importance of reflected incident
radiation and thermal emission can be estimated from the type of the primary,
the planetary Bond albedo, and the infrared flux.

A complete model of the reflected spectra of a cooling planet must
thus include all condensates and self-consistently account for the deposition of
incident radiation into the atmosphere.  Given the great range
in possible primaries and orbital distances, the dependence of Bond albedo 
upon the incident spectra, and the uncertainties inherent in cloud physics,
the overall phase space within which models must be constructed is
remarkably large.  

\acknowledgments
We thank Robert West for a thorough and helpul review and
K. Rages, T. Guillot, D. Saumon, and A. Burrows for helpful discussions.
The authors acknowledge the support of the 
NASA Faculty Awards for Research (MM), Planetary
Atmospheres (MM, RF), and Origins of Solar Systems (JL) programs 
as well as the NSF CAREER (MM) program.

\clearpage
\newpage
\figcaption{Geometric albedo spectrum for Jupiter and Uranus (Karkoschka 1994), 
compared to model spectra for a Jupiter-mass, solar-composition, planet
with Jupiter's effective temperature (128 K).  The model spectra
demonstrate the importance of clouds in controlling the reflected spectra.
Both models have a solar abundance of methane and no water.  
The cloudy model includes a stratospheric haze at 0.001 bar
(with an optical depth $\tau = 0.1$) and a tropospheric cloud at
1 bar ($\tau = 5$).  Both condensate layers scatter
conservatively and are gray.  The characteristics of Jupiter's clouds
have been derived by varying the locations and optical properties
to best fit the observed spectrum (e.g. Baines et al. 1989).
\label{fig1}}
\figcaption{Model phase integral spectra for a $12\,\rm M_J$
object with $T_{\rm eff} = 300 \,\rm K$. Phase integral is primarily
sensitive to scattering  properties of clouds, which in turn
depend upon particle size and composition. \label{fig2}}
\figcaption{(a) Water condensation and cloud size
profile for a warm, Jupiter-mass planet with $T_{\rm eff}=300\rm\, K$ 
and $g=22\,\rm m\,sec^{-2}$. Shown are the model temperature profile and 
the condensed water vapor density.  These quantities do not depend upon the
cloud model.  The particle size profile is shown for both a turbulent
and a quiescent cloud. (b) Same as (a), but for silicate ($\rm MgSiO_3$)
cloud and a hotter
(Gliese 229 B - like) model with $T_{\rm eff} = 1000\,\rm K$ and
$g =1000\,\rm m\,\sec^{-2}$. \label{fig3} }
\figcaption{Geometric albedo spectra for three models
with $T_{\rm eff}=128\,\rm K$ and varying gravity. 
All three atmospheres are quite dark in reflected light
beyond about $0.6\,\rm\mu m$.  Note log geometric albedo scale.  \label{fig4}}
\figcaption{Geometric albedo spectra for two model atmospheres,
each with Jupiter's gravity, but differing  effective
temperatures.   The Jupiter-like model differs from that shown in Figure 1
by having a solar abundance of water below the condensation level.
An ammonia band is apparent in the colder model near
$0.62\,\rm \mu m$.  \label{fig5}}
\figcaption{Geometric albedo spectra for cloud-free and water-cloud
models with $g=22\,\rm m\,\sec^{-2}$ and
$T_{\rm eff} = 300\,\rm K$.  (a)  Spectra of atmospheres with no cloud,
turbulent cloud, and quiescent cloud.  To clearly demonstrate the influence
of the cloud model upon the spectra, $f_s = 1$
is assumed.  Particle sizes are smaller
for the quiescent cloud which enhances the geometric albedo
at the shortest wavelengths. In the near-infrared the greater
column abundance of the smaller particles enhances scattering
substantially for the radiative cloud.  Both cloud models
are much brighter than an atmosphere with no clouds.
(b) Comparison of spectra assuming clouds with various
supersaturation factors.  For $f_s=0.01$ effect of cloud
is only apparent beyond about $1\,\rm \mu m$ where the 100-fold
decrease in cloud particle column density translates
into about a 10-fold drop in albedo which is nevertheless
far brighter than for a cloud-free atmosphere (Figure 4).
(c) Spectra for atmospheres with single particle size clouds,
all with $f_s=0.01$.  Optical and near-IR scattering properties
are similar for all cases with $r_c\ge 10\,\rm \mu m$.
\label{fig6}}
\figcaption{Similar to Figure 6 but
for silicate clouds in the atmosphere of a brown dwarf
with $T_{\rm eff} = 1000\,\rm K.$ and $g=1000\,\rm m\,sec^{-2}$.
Model object, similar to $\rm  G\ell 229\, B$, has
a mass of about $36\,\rm M_J$.  The silicate clouds lower
the albedo in the blue but are less detectable in
the near-IR. (a) Comparison of cloud models with $f_s=1$.
(b) Effect of $f_s$ on spectra.  (c) Role of particle size. \label{fig7}}
\figcaption{Bond albedos for atmosphere with
$g=22\,\rm m\,sec^{-2}$  and $T_{\rm eff} = 300\,\rm K$
and various model clouds as a function 
of the primary type.  Symbols summarize cloud type, either
turbulent, quiescent, or single-sized. Also shown (slightly
offset for clarity) are
the Bond albedos and associated uncertainties for
Jupiter and Neptune (Conrath et al. 1989).  (a) Models with
$f_s=0.01$. (b) Models with $f_s=0.1$.  \label{fig8}}

\figcaption{Similar to Figures 8 but for 
model with $g=1000\,\rm m\,sec^{-2}$  and $T_{\rm eff} = 1000\,\rm K$
and various $\rm MgSiO_3$ clouds.  Note differing vertical scale
from Figures 8. \label {fig9}}
\figcaption{Composite spectra for 
model with $g=300\,\rm m\,sec^{-2}$  and $T_{\rm eff} = 300\,\rm K$.
Lines, as indicated,  show reflected spectra which would be measured
at 10 pc from a planet orbiting at 1 AU from its primary.  Long
dashed line shows thermal emission from the same atmosphere model 
with no clouds as computed by Burrows
et al. (1997).  Reflected light dominates thermal emission in near-infrared
for cloudy models with effective temperatures below this range.
A number of photometric bandpasses are shown for reference.
\label{fig10}}
\figcaption{As for Figure 10, but for a roughly $36\,\rm M_J$
model with $g=1000\,\rm m\,sec^{-2}$  and $T_{\rm eff} = 1000\,\rm K$
and silicate clouds.  Thermal emission dominates reflected flux 
for such hot models.  For a planet orbiting inside of $0.1\,\rm AU$,
reflected flux would be comparable to emitted flux.
\label{fig11}}

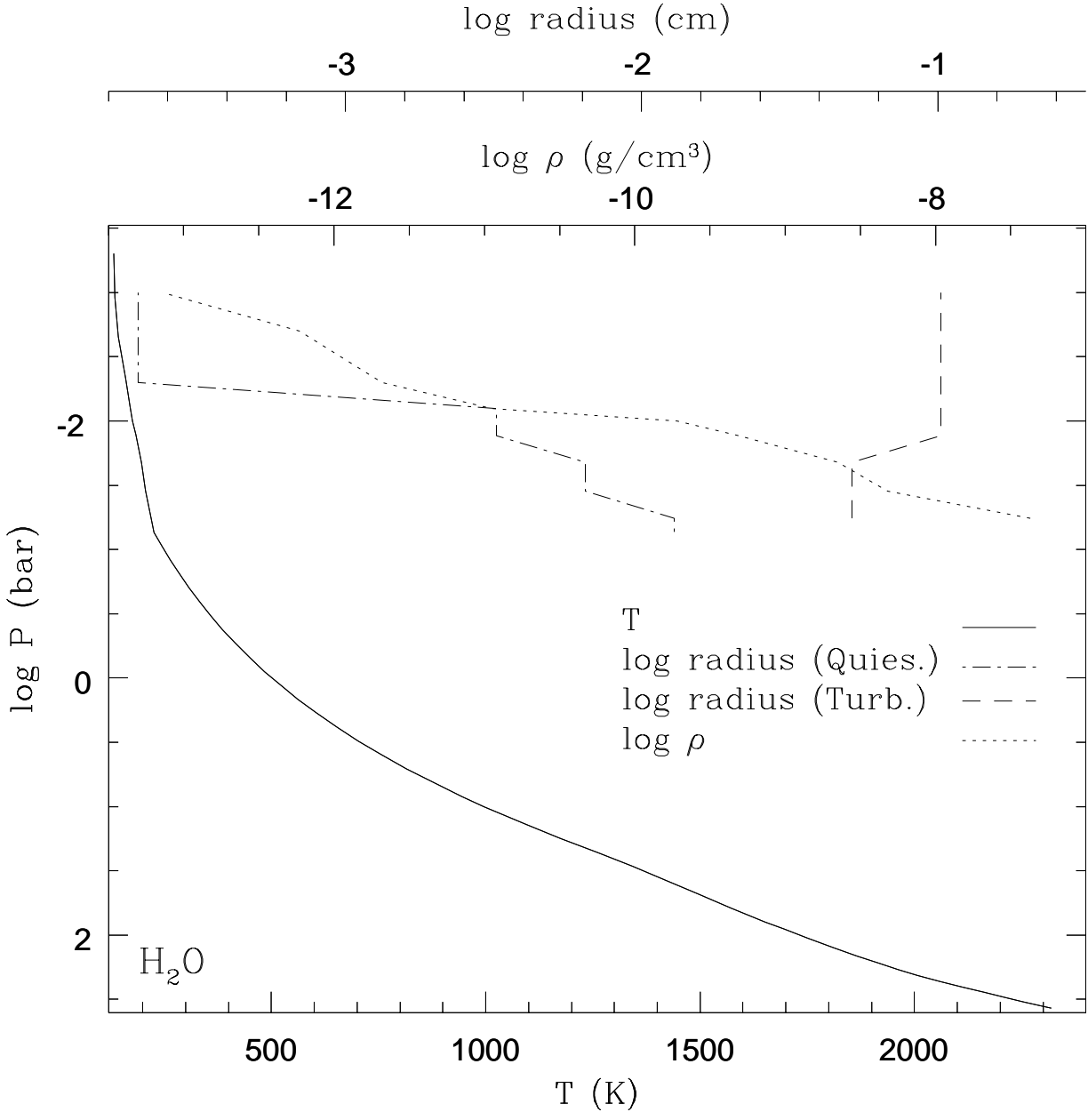
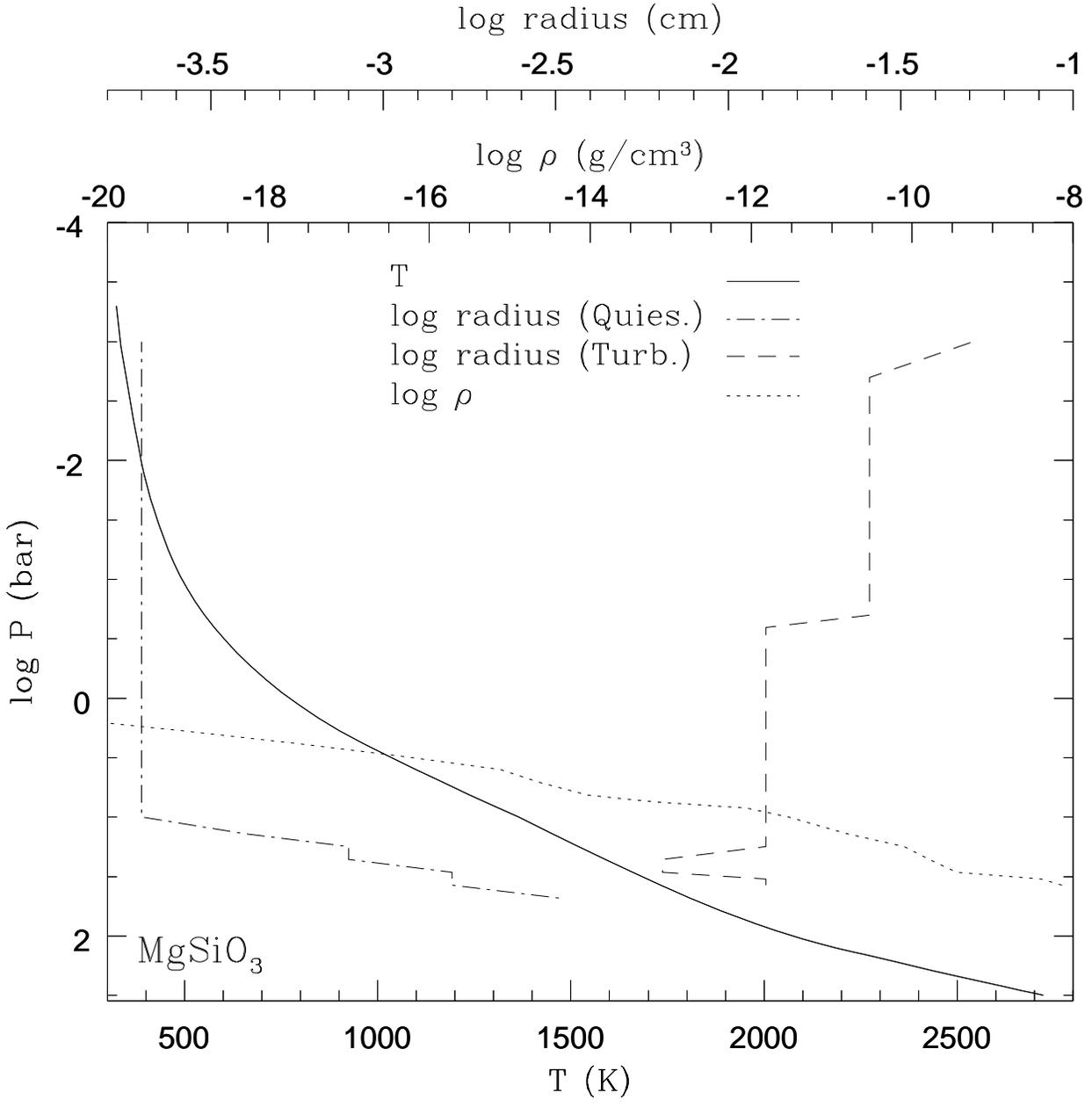
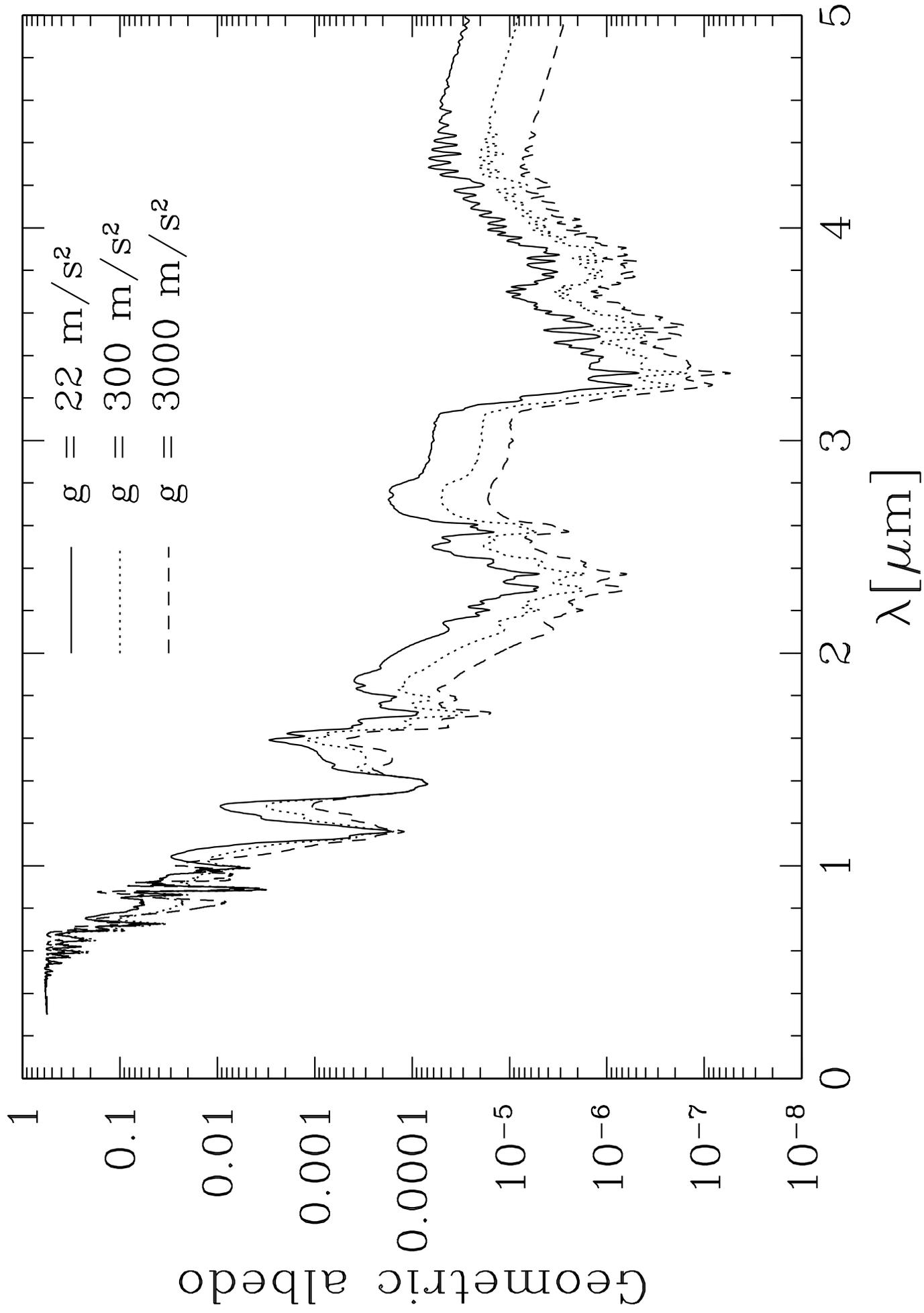
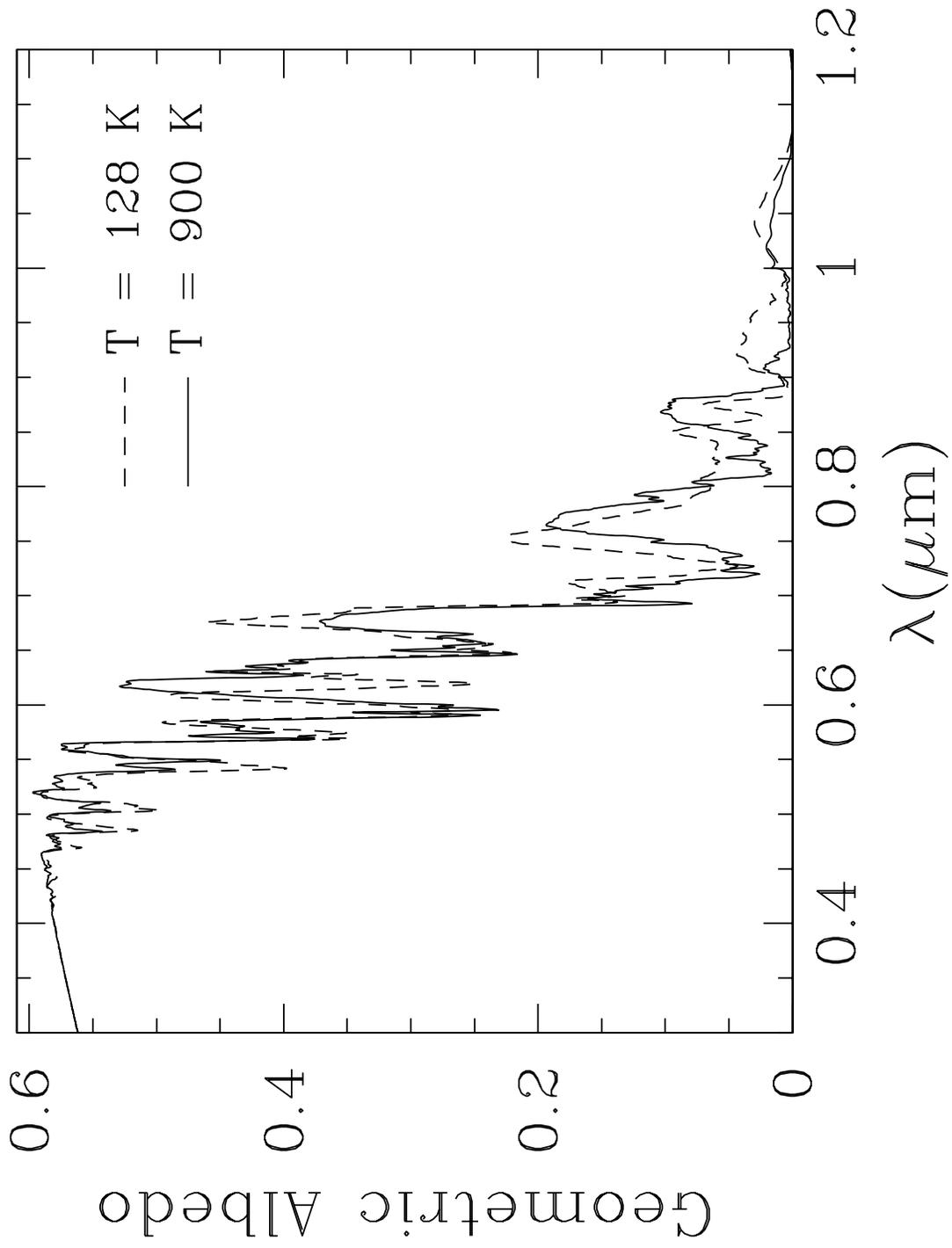
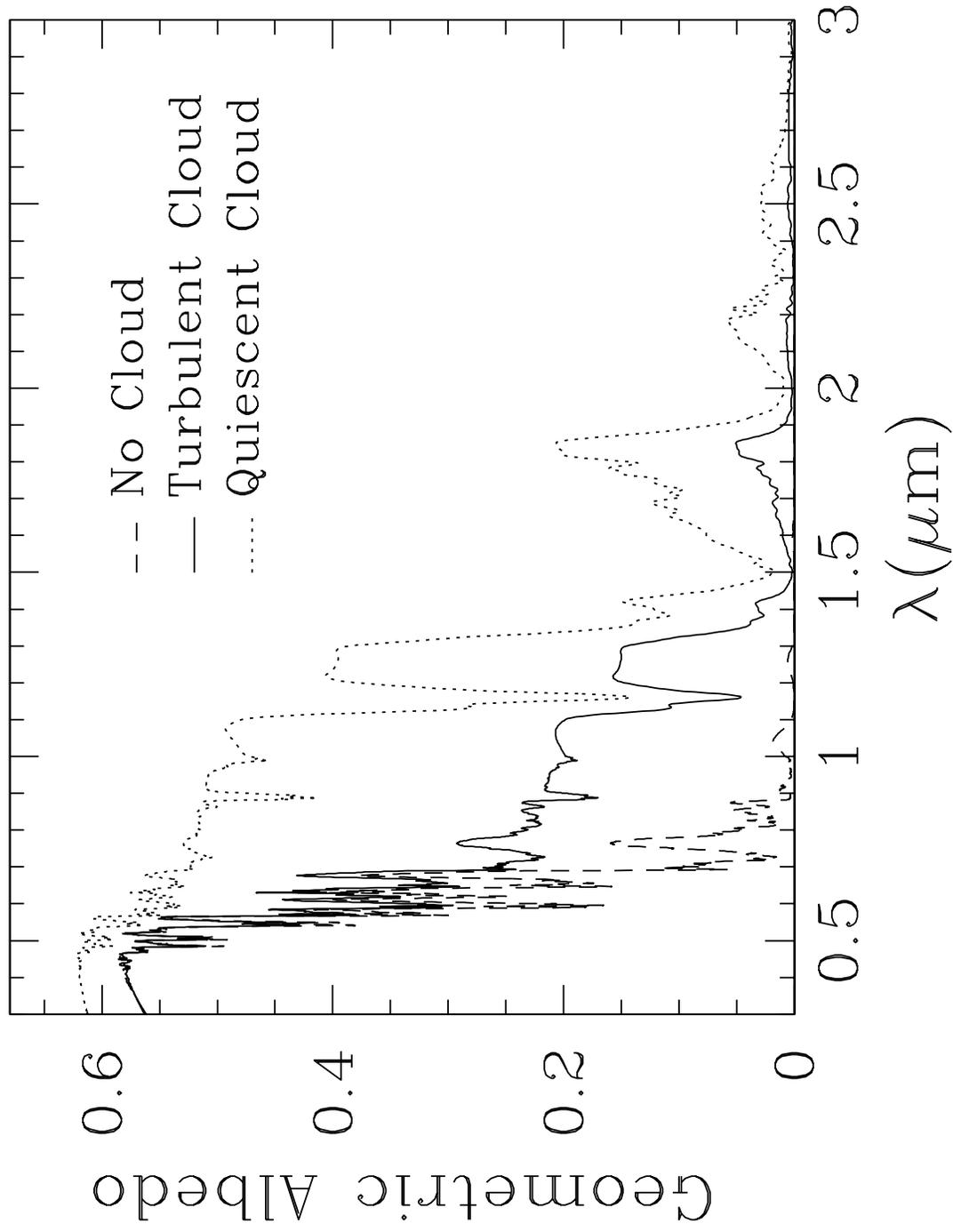
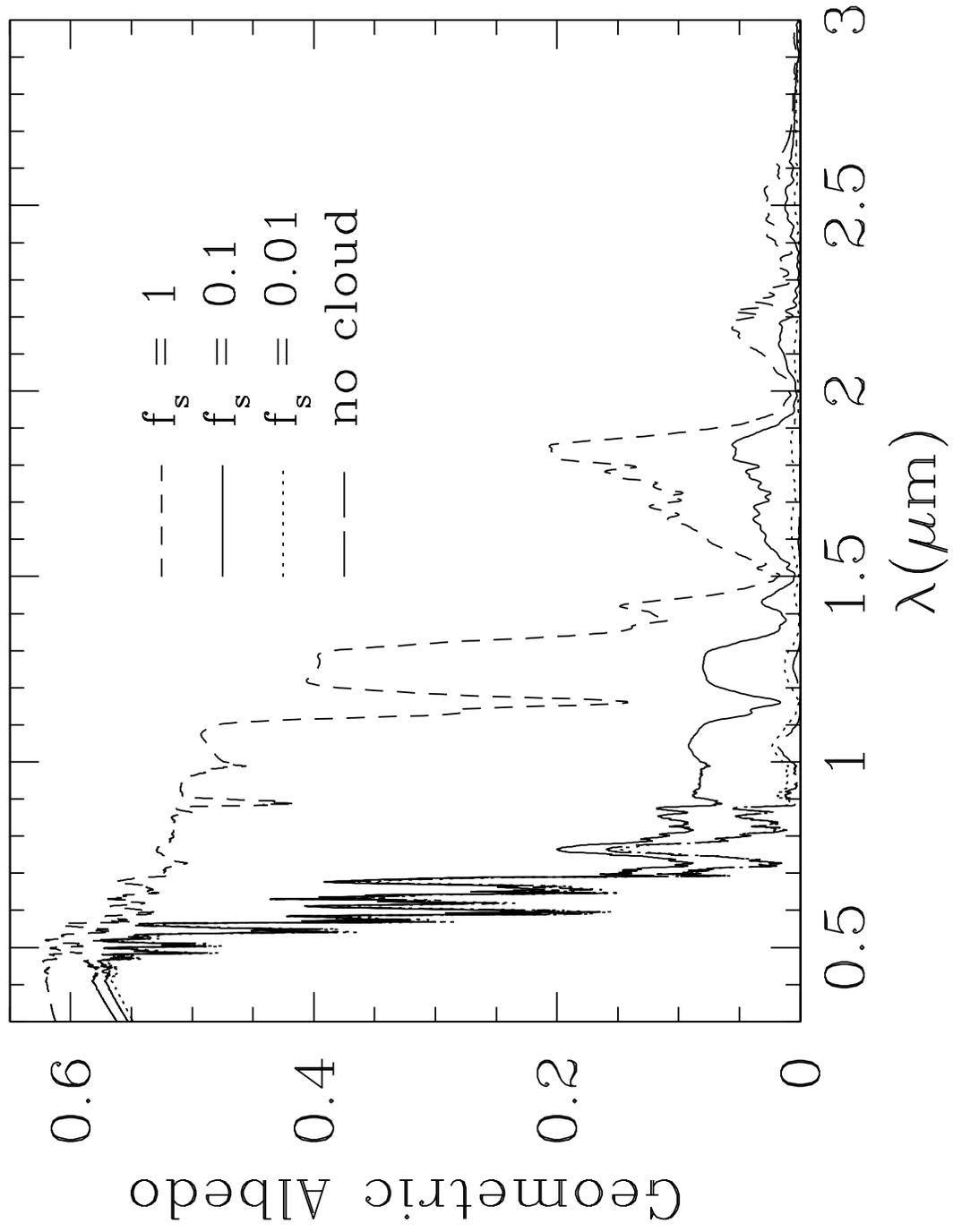
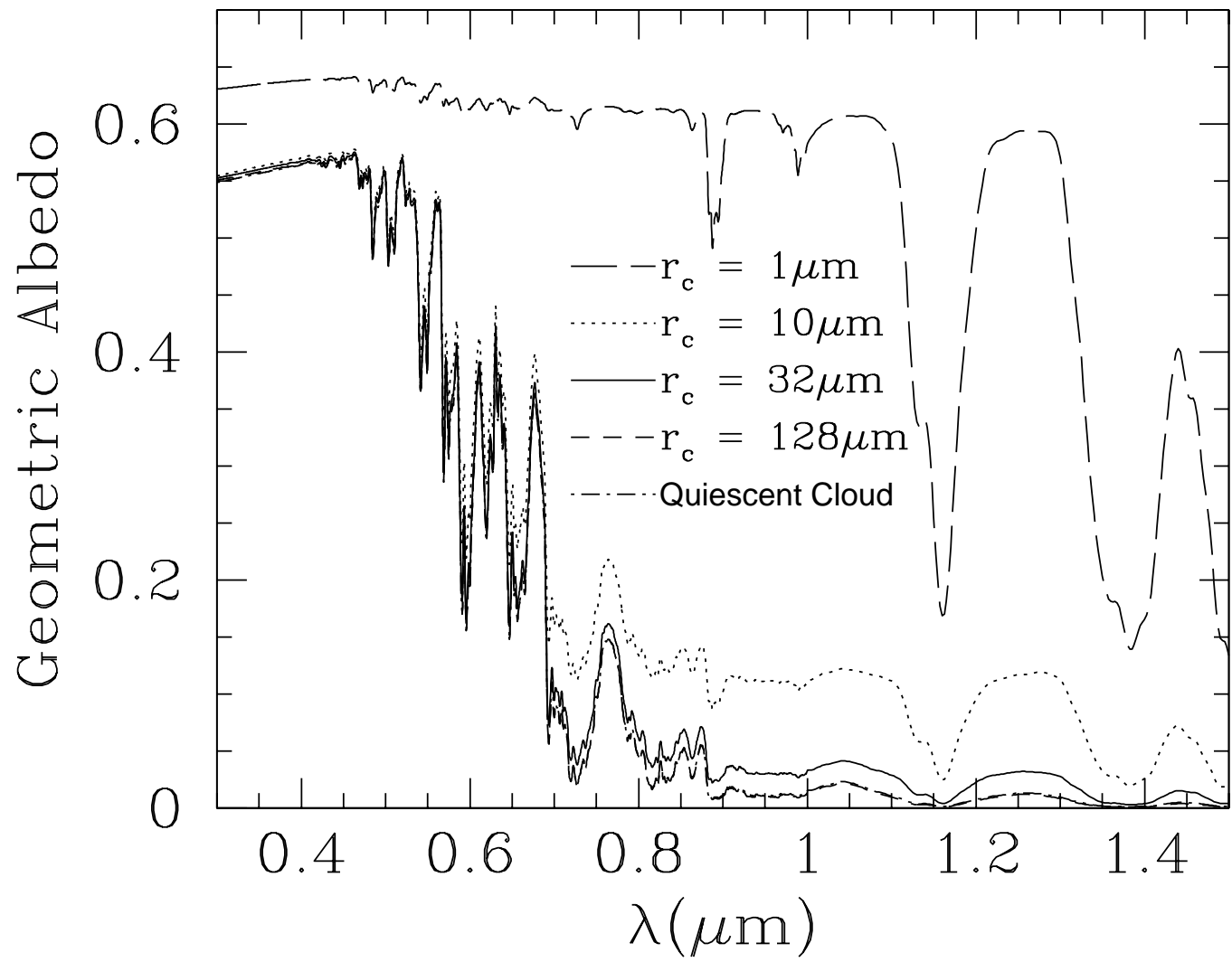
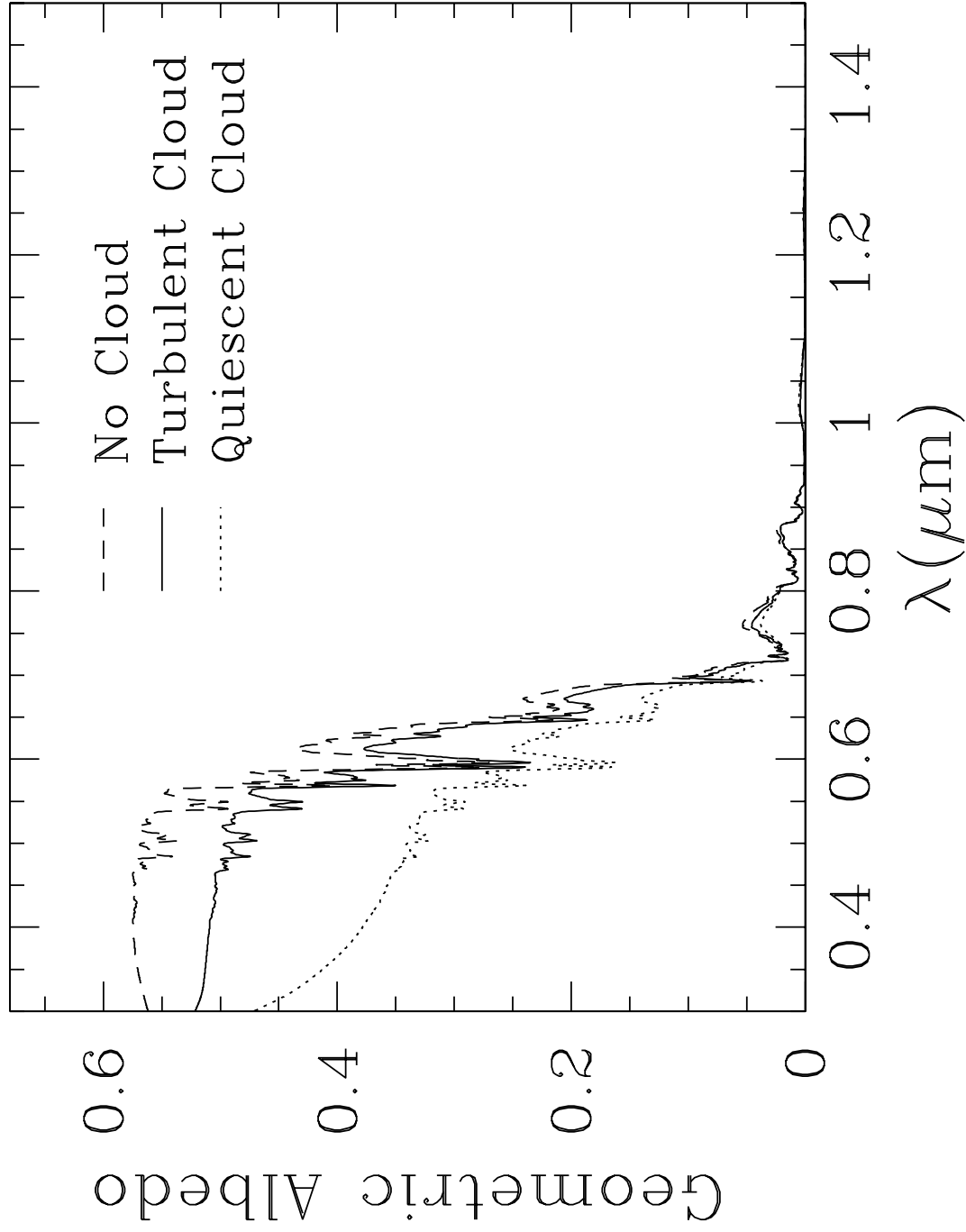
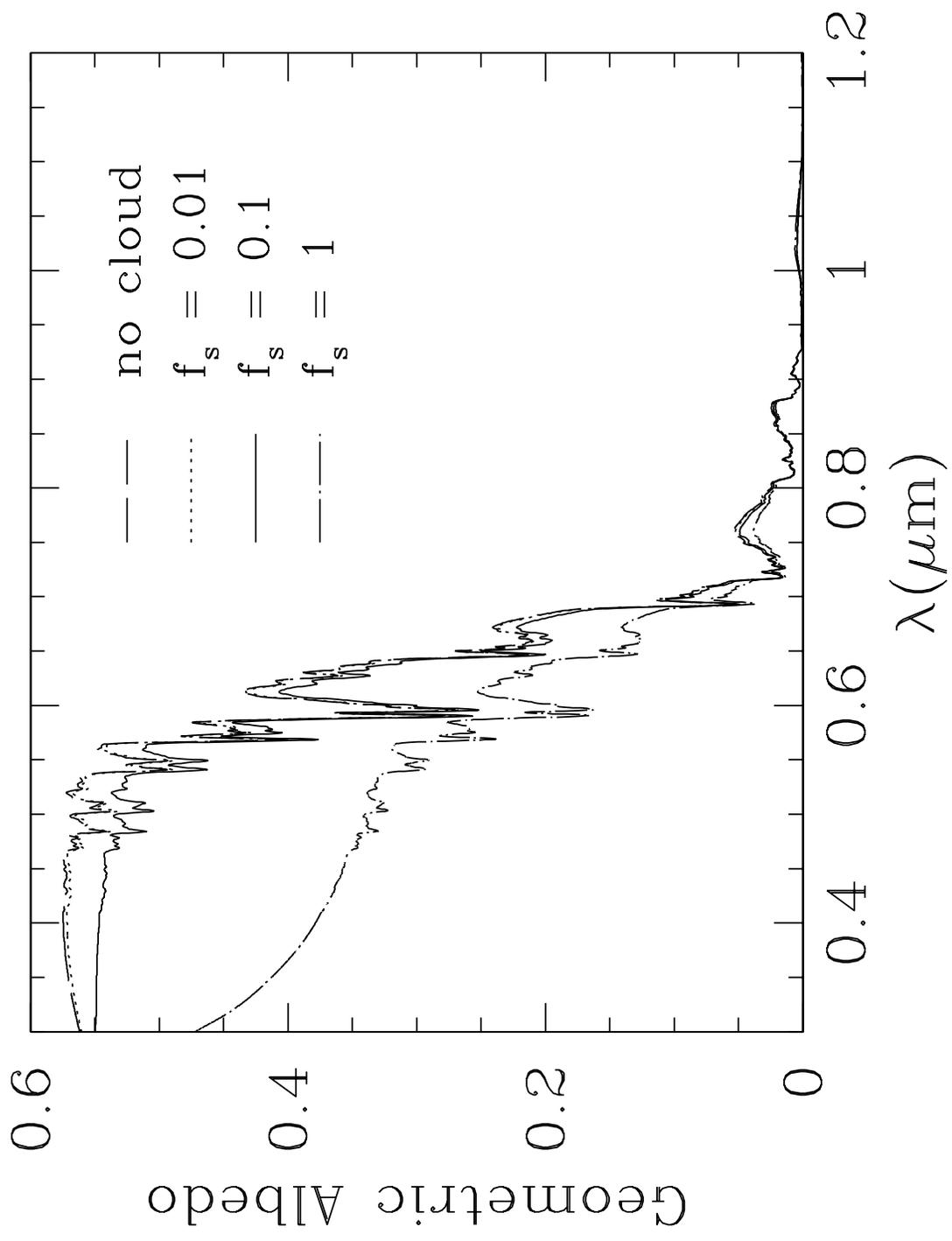
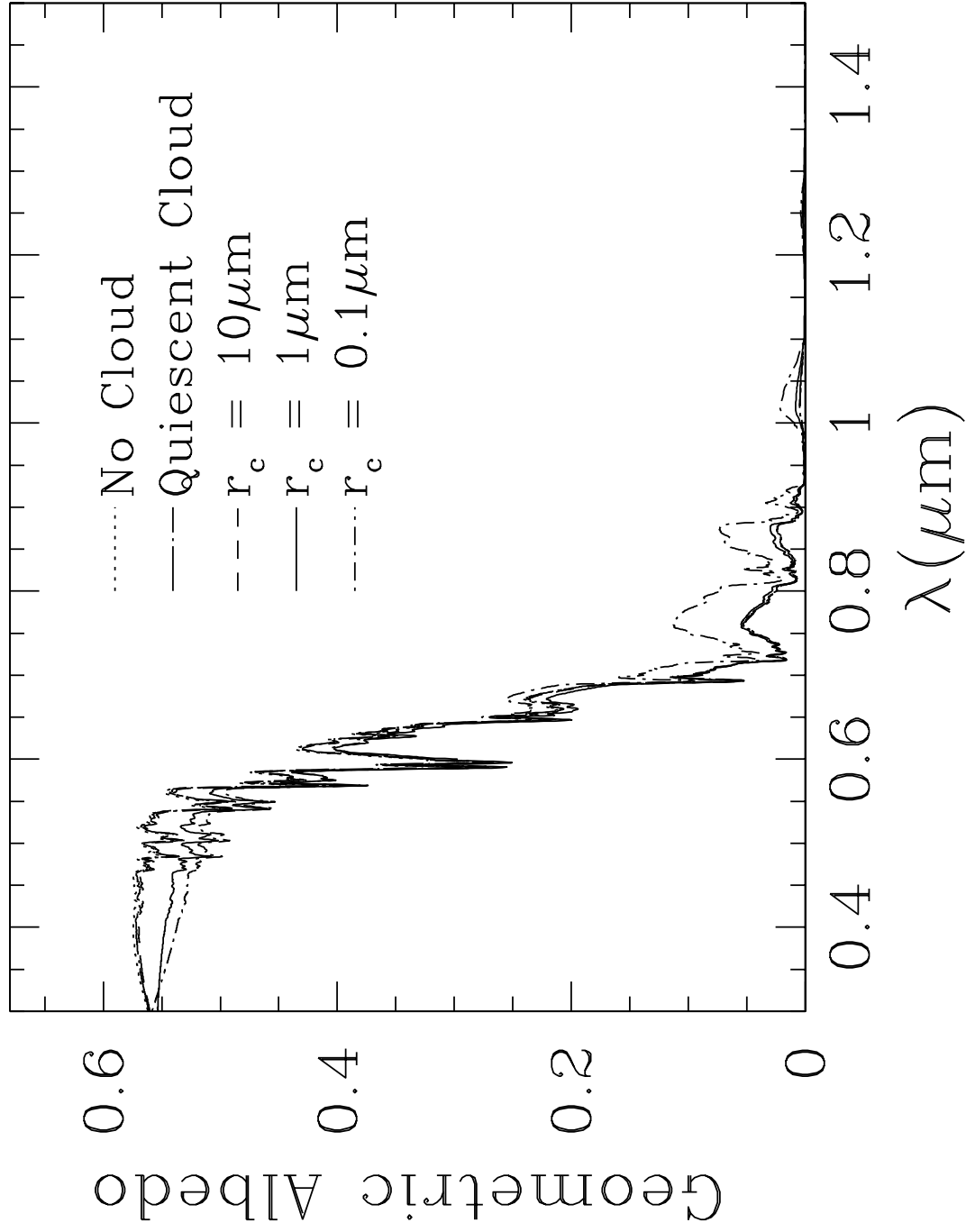
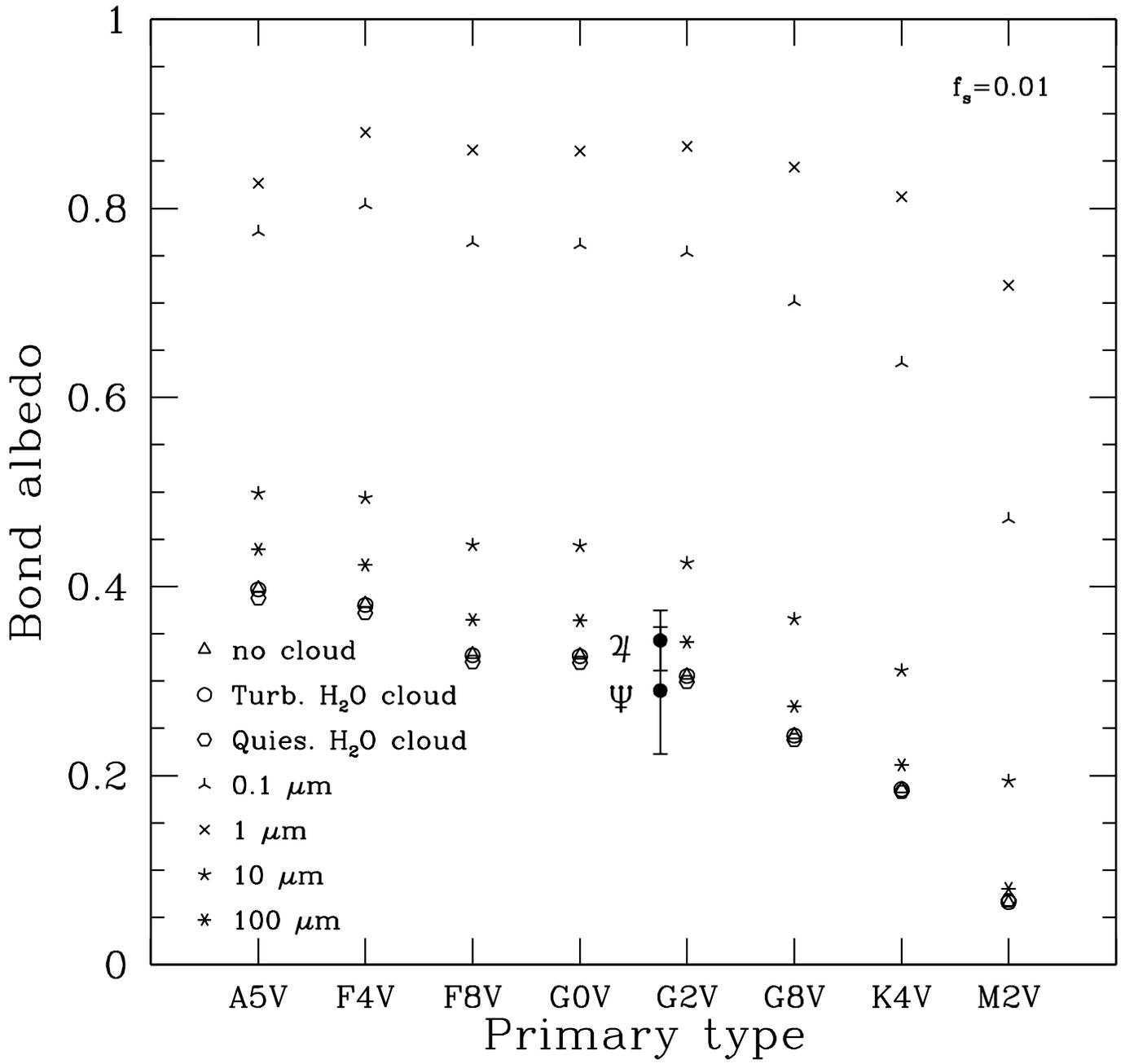
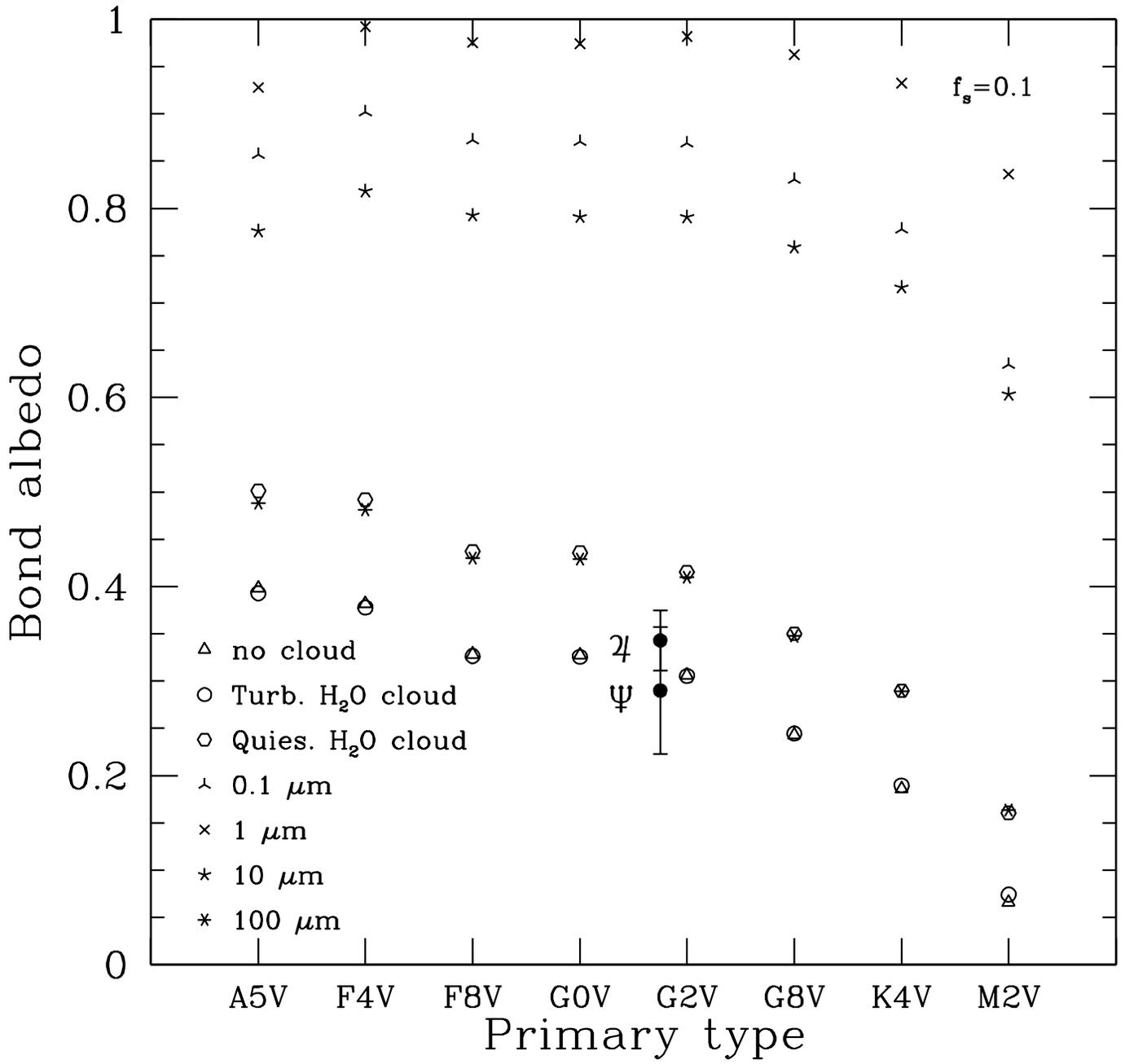
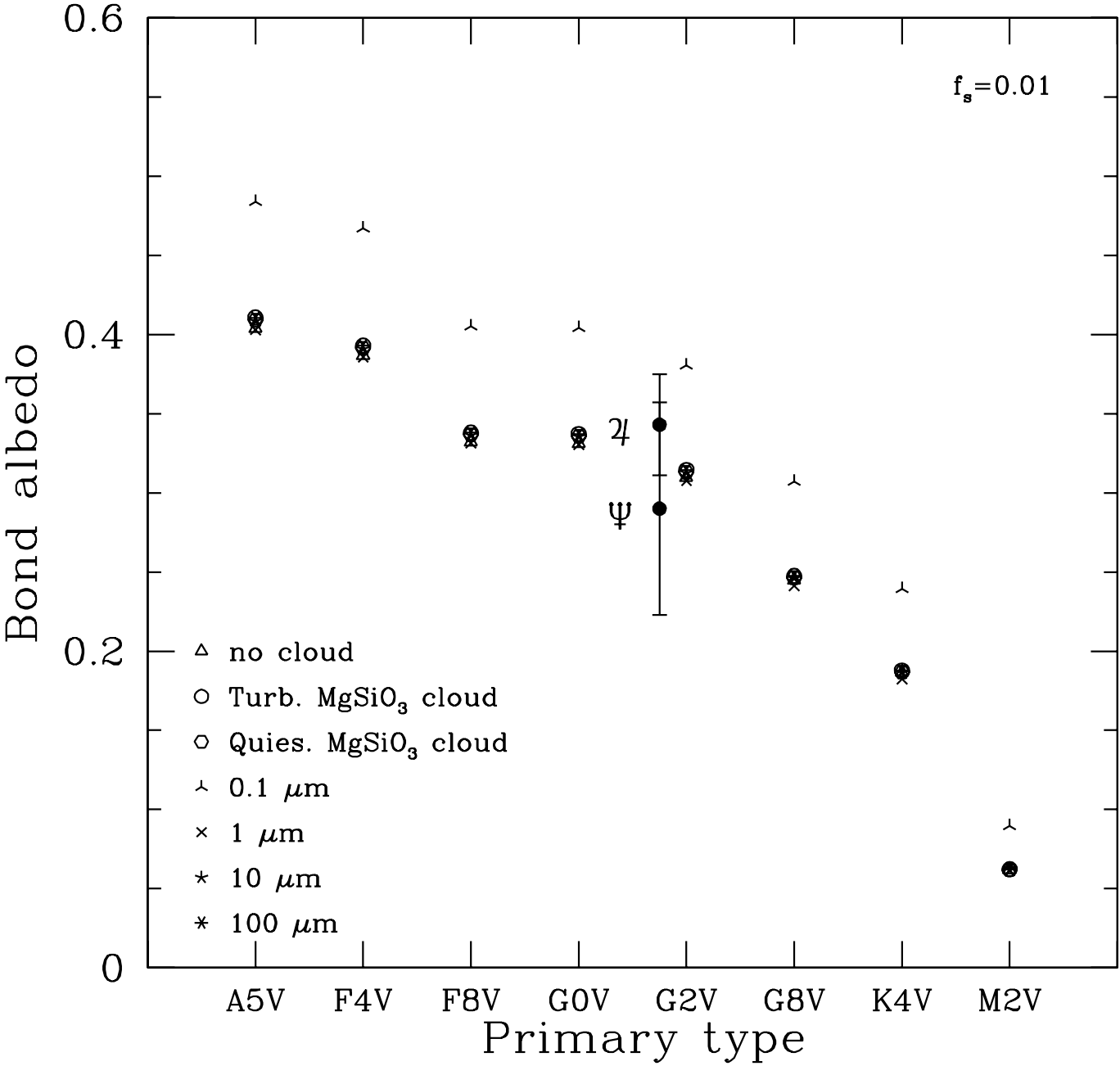
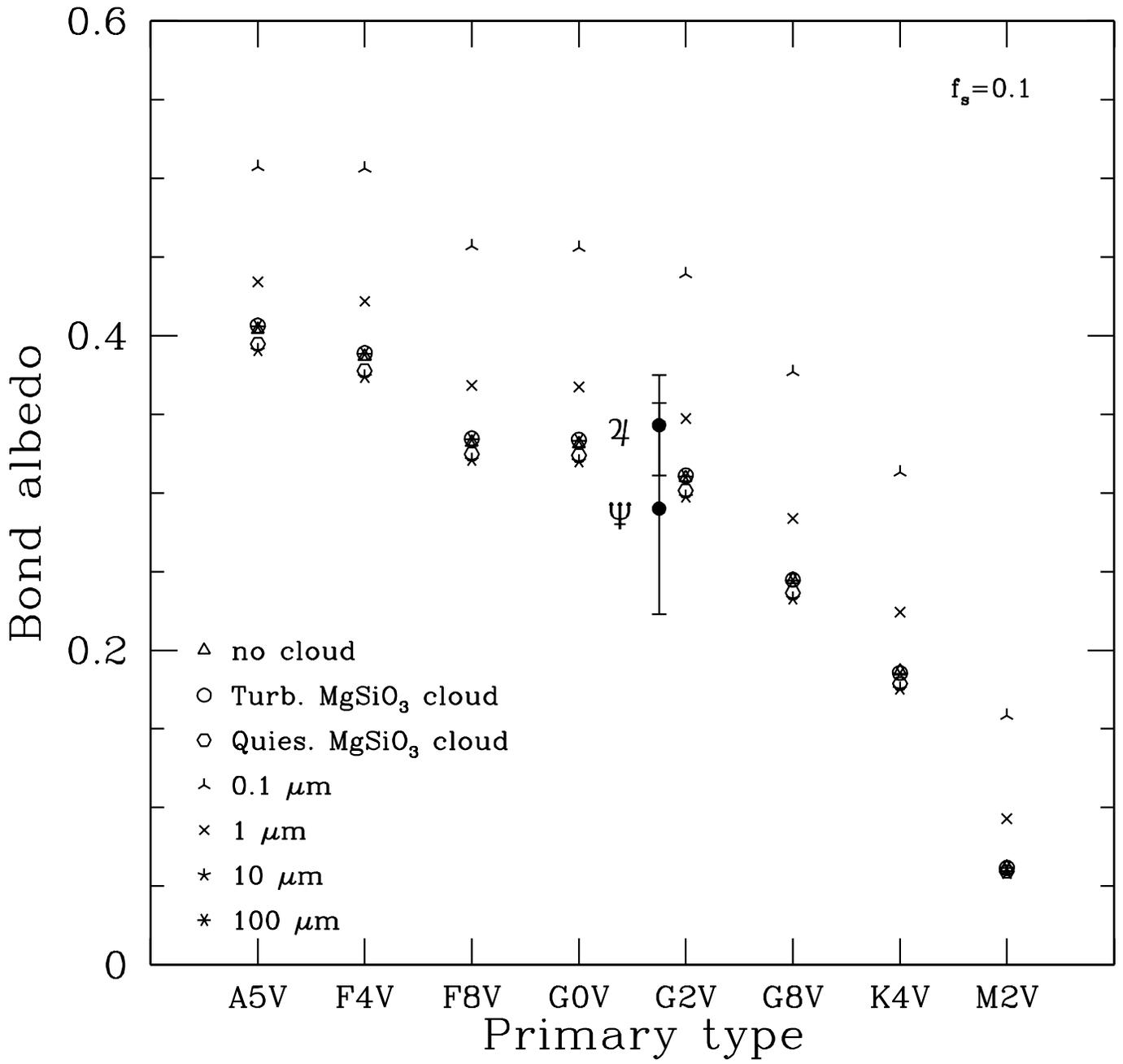
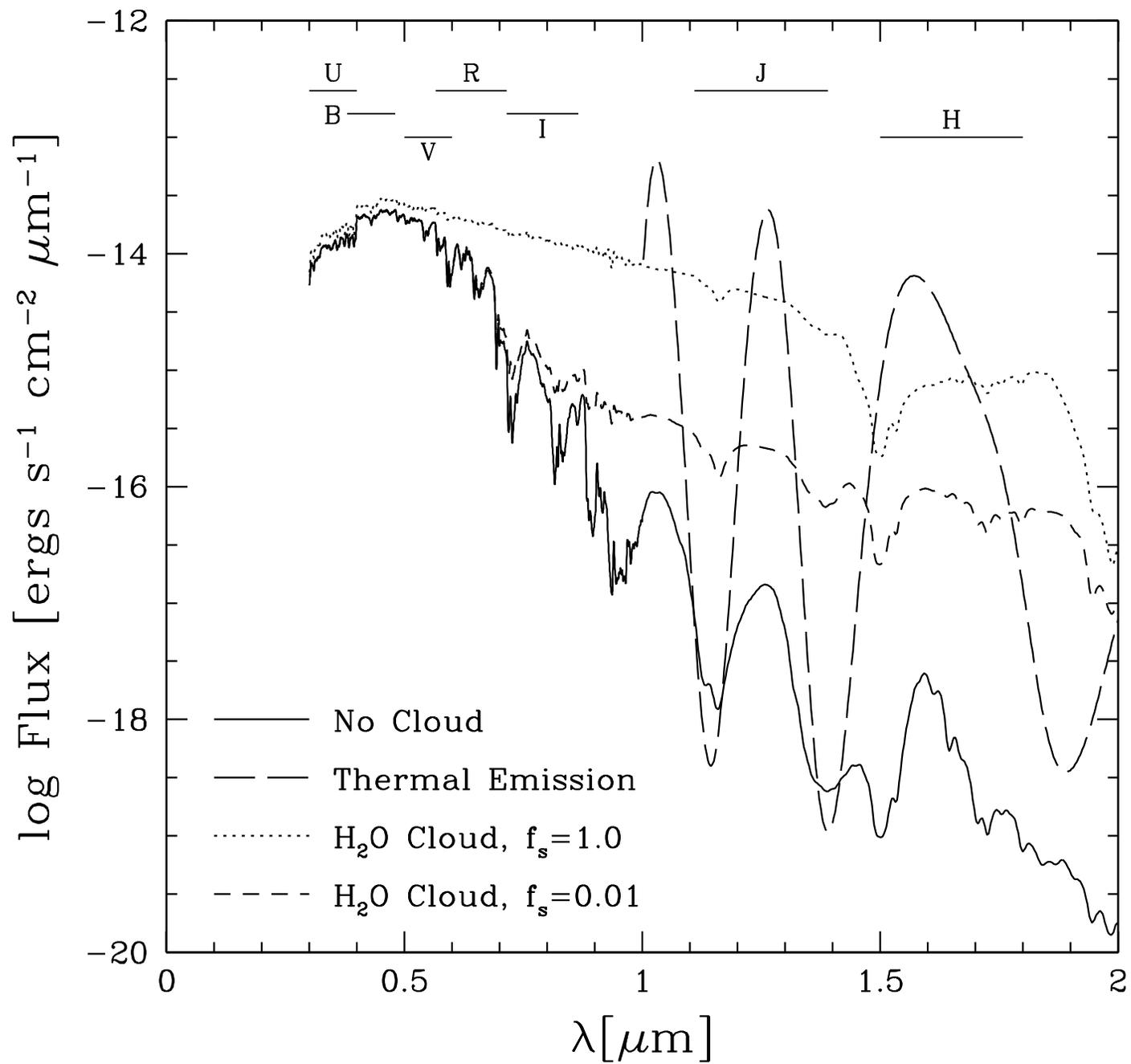
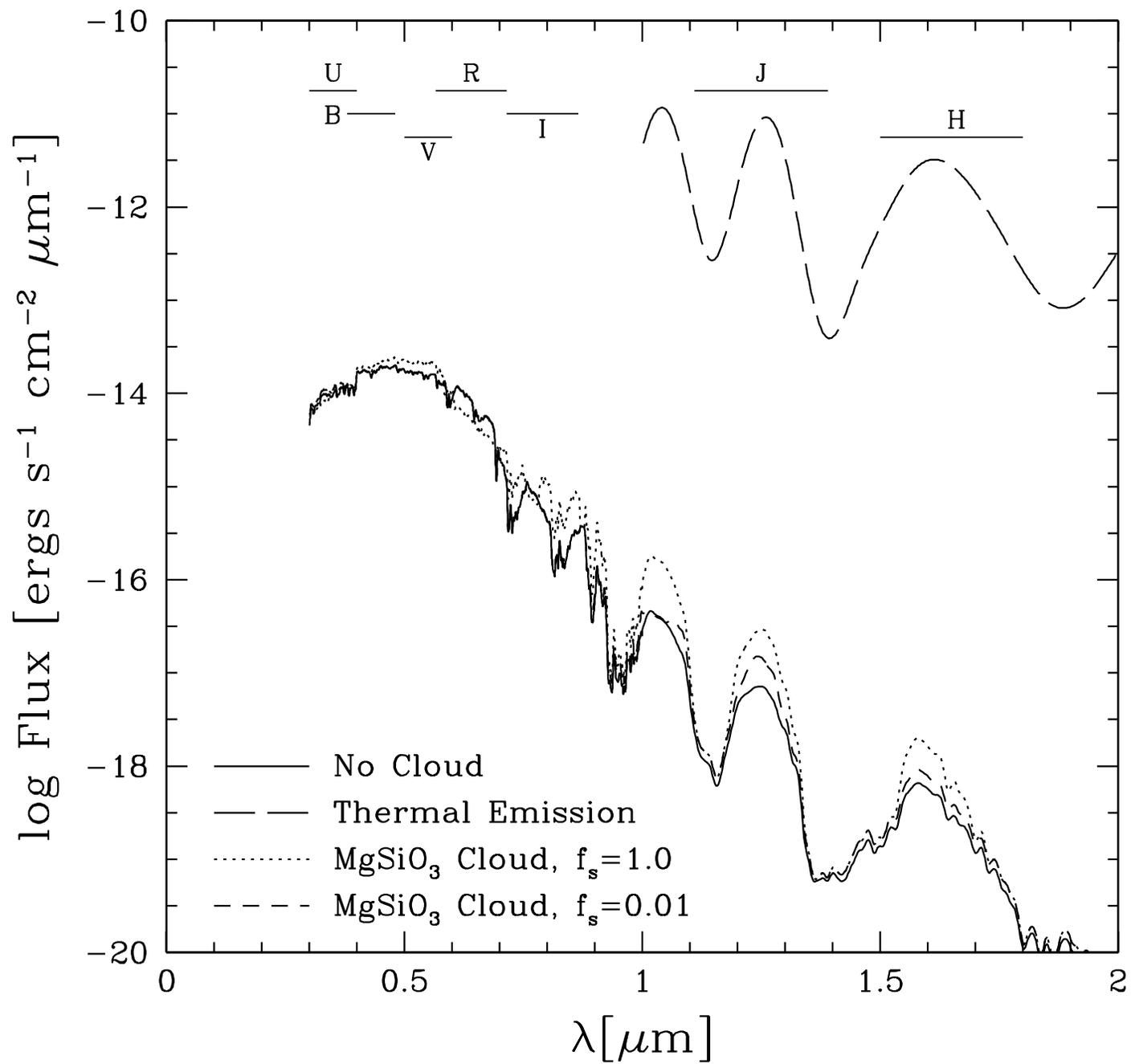
\end{document}